\documentclass[11pt,twoside]{JHEP3}
\usepackage{graphicx}
\usepackage{amsmath}
\usepackage{amssymb}

\def\be{\begin{equation}}
\def\ee{\end{equation}}
\def\bse{\begin{subequations}}
\def\ese{\end{subequations}}
\def\ba{\begin{array}}
\def\ea{\end{array}}
\def\bea{\begin{eqnarray}}
\def\eea{\end{eqnarray}}
\def\pr{\partial}
\def\nno{\nonumber}

\def\vth{\vartheta}
\def\vph{\varphi}

\title {Charged black holes in generalized dilaton-axion gravity}

\author{Sourav Sur\\
Institute of Physics\\
Bhubaneswar - 751 005, India\\
e-mail: \email{sourav@iopb.res.in}}

\author{Saurya Das\\
Department of Physics\\
University of Lethbridge\\
4401 University Drive, Lethbridge, Alberta T1K 3M4, Canada\\
e-mail: \email{saurya.das@uleth.ca}}

\author{Soumitra SenGupta\\
Department of Theoretical Physics\\
Indian Association for the Cultivation of Science\\
Jadavpur, Kolkata - 700 032, India\\
e-mail: \email{tpssg@iacs.res.in}}

\abstract{We study generic Einstein-Maxwell-Kalb-Ramond-dilaton actions,
and derive conditions under which they give rise to static, 
spherically symmetric black hole solutions. We obtain new asymptotically 
flat and non-flat black hole solutions which are in general electrically 
and magnetically charged. They have positive definite and finite quasi-local 
masses. Existing non-rotating black hole solutions (including those 
appearing in low energy string theory) are recovered in special limits. 
}

\begin{document}
\section{Introduction   \label{intro}}
Theories of gravity, with scalar and/or pseudoscalar
field(s) in the background, have been under investigation for a long time. 
Such theories, taken in conjunction with the couplings of the
(pseudo)scalar field(s) with vector or tensor gauge fields, turn out to be 
very promising to obtain various black hole solutions
of the gravitational field equations. For gravity
coupled minimally to one or more scalar field(s), it has long
been established \cite{bek}, as an immediate follow-up of the
`no hair' theorem, that a black hole cannot support the additional
scalar degree(s) of freedom in a static spherically symmetric
spacetime which is required to be asymptotically flat. This is
true for all scalar fields, and evidently for all pseudoscalar
fields as well, regardless of their mass or any positive (convex) potential
assigned to them. Attempts have therefore been made either to
explore non-minimal (conformal) scalar couplings with gravity
\cite{conform}, or to invoke a non-vanishing 
cosmological constant within the usual minimal coupling prescription
\cite{cc}, in order to obtain regular and stable black hole solutions
of the effective Einstein's equations. 
It has further been shown that in  presence of 
both gauge fields as well as  scalar/pseudoscalar fields in a 
curved background, black hole solutions exist for
some specific choices of the couplings between the 
(pseudo)scalar(s) and the gauge fields.
Such couplings arise naturally in some string theories
and are responsible for the existence of black holes in such theories.
There have been extensive investigations for black holes in
string theory (see \cite{strrev} for some comprehensive reviews) and
the implications of the typical (pseudo)scalar--gauge field couplings
(which are intimately associated with the duality relationships) have
been studied in detail both in asymptotically flat (AF) spacetimes
\cite{gib}-\cite{all} as well as in asymptotically non-flat (AN) 
spacetimes \cite{mann}-\cite{clement}. These black hole solutions, 
characterized by the gauge charges, can be useful in 
exploring the effects of string theory in the low energy world.

In this paper we examine the possibility of the 
existence of charged black holes in a theory of gravity coupled with
scalar/pseudoscalar fields and a $U (1)$ (electromagnetic) gauge
field. Our approach is to start with arbitrary scalar
and pseudoscalar couplings with the invariant quantities $F_{\mu\nu}
F^{\mu\nu}$ and $F_{\mu\nu} ~^\star F^{\mu\nu}$, in the action 
(where $F_{\mu\nu}$ is the
electromagnetic field strength and $^\star F^{\mu\nu}$ is its dual). 
From such a general action we try
to construct charged black hole solutions of the
gravitational field equations in four dimensions.
Requiring the existence of a horizon, 
we determine the nature of the scalar-electromagnetic couplings. 
This in a way generalizes the well-known scalar-electromagnetic coupling 
in string theory leading to black hole solutions. 

As is well known \cite{gsw,pol}, the massless spectrum of the 
bosonic sector of string theory consists of the 
graviton, the vector gauge
fields, the second rank antisymmetric tensor 
Kalb-Ramond (KR) field \cite{kr} and
the scalar dilaton. The dilaton has an exponential coupling 
with all the other fields in the action. The four dimensional effective action 
a theory of gravity, derived from string theory, is thus the  standard 
Einstein-Hilbert action coupled to  the dilaton $\phi$,
the KR field $B_{\mu\nu}$ and a set of $U (1)$ gauge fields $A_{\mu}^i~
(i = 1, \cdots, N)$. For simplicity we consider only one $U(1)$ gauge field
$A_\mu$ --- the Maxwell (electromagnetic) field. The KR
field strength is, however, augmented with the $U (1)$ Chern-Simons term
in order to keep the corresponding quantum theory free from gauge
anomaly\footnote{There is an additional augmentation of the KR field strength
with the Lorentz Chern-Simons term on account of gravitational anomaly
cancellation. However, we are not taking any notice of it since it is
dimensionally suppressed, for being quadratic in the scalar curvature.
In fact, we are not taking into account the higher curvature corrections
to the tree level string action anyway.}. The Chern-Simons term is found
to play a crucial role in preserving gauge symmetry under both $U (1)$ and
KR gauge transformations \cite{pmssg}.
Further, the Chern-Simons augmented third rank field strength corresponding to
the second rank KR field can be expressed as a hodge dual of 
the derivative of a  
pseudoscalar field (called axion $\xi$) in four dimensions.

The gauge-invariant KR-electromagnetic coupling, which appears as a $\xi
F ~^\star F$ term in the effective action, has an interesting 
astrophysical/cosmologicalimplication, namely cosmic optical birefringence. 
This has been studied in detail in some recent works both in four dimensions 
\cite{optact} and also in the context of higher dimensional Randall-Sundrum 
\cite{rs} scenario \cite{dmssg,dmssgssnew,acpm}. The effect of the KR field on 
various other physical phenomena, such as gravitational lensing and redshift 
\cite{skssgss}, neutrino oscillation \cite{ssgas}, Cosmic Microwave polarization 
anisotropy \cite{dmpmssg}, cosmic acceleration \cite{ssgsscosm} etc. have also 
been explored to look for possible indirect evidences of string theory at low
energies. On the other hand, the effects of the dilaton are of substantial
interest both in cosmology (see \cite{dilcosm} for some reviews) as well in the
context of charged black holes and their implications in string theory 
\cite{strrev}.

In this paper we focus on charged black hole solutions in a KR-dilaton background.
We resort to a string inspired scenario in which the standard Einstein-Maxwell 
theory is supplemented by two scalar/pseudoscalar fields $\phi$ and $\xi$, which 
we later identify as the stringy dilaton and axion fields in four dimensions. We 
consider generic couplings (as arbitrary functions of $\phi$ and $\xi$) between 
these (pseudo)scalars and the electromagnetic field and also take into account an 
arbitrary interaction between $\phi$ and $\xi$. With this setup, in section 
\ref{general}, we arrive at an effective action that involves a single scalar 
field $\psi$ (which is a function of $\phi$ and $\xi$) coupled to the 
Einstein-Maxwell system. Resorting to a generic ansatz for the metric, we obtain 
exact solutions of the gravitational field equations, representing static 
spherically symmetric charged black holes. In the process, the general forms of 
the effective scalar $\psi$ and its coupling with the electromagnetic field is 
determined in terms of the radial variable $r$. It turns out that the asymptotic 
properties of solutions (AF or AN) depends upon some free parameters present in 
our ansatz. For both AF and AN types of spacetimes, we work out the general 
expressions for the gravitating mass $m ~(\equiv G M)$ following the standard 
quasi-local mass formalisms \cite{brown,hawkhor} in section \ref{massform}. In 
section \ref{gensol}, we obtain the solution for the metric of both AF and AN 
types and that for the effective scalar field $\psi$. Finally, in section 
\ref{dilax} we analyze our solutions from the point of view of the low energy 
effective string theory (as well as some possible generalizations of this) by 
varying the parameters of the action. The possible origins of these parameters 
are explained in section \ref{dilax}. We work out the complete set of solutions 
separately for both AF and AN spacetimes for some specific values of these 
parameters. In this process we reproduce some known solutions in dilaton-axion 
gravity as well as find some new ones.

It is also to be mentioned here that in most of this paper, we work in the Einstein 
frame, in which the scalars are minimally coupled to gravity and the Planck length, 
which is proportional to square root of the universal gravitational constant $G$, 
is taken as the fundamental length scale. In dealing with the string background 
geometries, it is in fact always convenient to work in the Einstein frame, rather 
than the string frame which is related to the Einstein frame via the conformal 
transformation\footnote{In our convention $\phi$ is the dilaton, rather than $\Phi 
= \phi/2$ as mostly in the literature.} $g_{\mu\nu} = e^{- \phi} \tilde{g}_{\mu\nu}$, 
where $g_{\mu\nu}$ and $\tilde{g}_{\mu\nu}$ are respectively the Einstein metric and 
the string metric. In the string frame, the string length scale $l_s$ is taken as 
the fundamental unit for all measurements, while the effective Planck scale varies 
with the dilaton. As such, there is no usual way of a physical realization of any 
experimental data in such a frame.

\section{General Formalism \label{general}}

Let us begin with a string inspired scalar coupled Einstein-Hilbert-Maxwell
action in four dimensions (in Einstein frame)
\be \label{gen-action}
S ~=~ \frac 1 {2 \kappa} \int d^4 x \sqrt{-g} \left[R - \frac 1 2 \pr_{\mu} \phi
\pr^{\mu} \phi - \frac{\omega (\phi)} 2 \pr_{\mu} \xi \pr^{\mu} \xi -
\alpha (\phi, \xi) F_{\mu\nu} F^{\mu\nu} - \beta (\phi, \xi) F_{\mu\nu}
~^\star F^{\mu\nu}\right].
\ee
where $\kappa = 8 \pi G$ is the four dimensional gravitational coupling
constant. $R$ is the curvature scalar, $F (= \pr_{[\mu} A_{\nu]})$ is the
field strength of the Maxwell field $A_{\nu}$, and $\{\phi,\xi\}$ are two
massless scalar/pseudoscalar fields where $\xi$ acquires a non-minimal 
kinetic term given by the function $\omega (\phi)$ because of it's 
interaction with the scalar $\phi$. In addition, the functions $\alpha$ 
and $\beta$ describe the couplings of $\phi$ and $\xi$ with the Maxwell 
field which at this stage are chosen to be as general as possible. The 
free Maxwell action consists of both the scalar and pseudoscalar invariants, 
$F_{\mu\nu} F^{\mu\nu}$ and $F_{\mu\nu} ~^\star F^{\mu\nu}$, where ~$^\star F^{\mu\nu} 
= \frac 1 2 \epsilon^{\mu\nu\alpha\beta} F_{\alpha\beta}$ is the (Hodge-) 
dual Maxwell field strength.

The field equations corresponding to the above action (\ref{gen-action}) are
given by
\bse
\label{gen-feq}
\bea
&&R_{\mu\nu} ~=~ \frac 1 2 \pr_{\mu} \phi \pr_{\nu} \phi ~+~ \frac{\omega (\phi)} 2
\pr_{\mu} \xi \pr_{\nu} \xi ~+~ 2 \alpha (\phi, \xi) \left(F_{\mu\rho} F_{\nu}^{~\rho} ~-~
\frac 1 4 g_{\mu\nu} F_{\rho\lambda} F^{\rho\lambda}\right)  \label{gen-feq-a} \\
&&\nabla_{\mu} \pr^{\mu} \phi ~=~ \frac 1 2 \pr_{\phi} \omega ~\pr_{\mu} \xi\pr^{\mu} \xi 
~+~ \pr_{\phi} \alpha ~F_{\mu\nu} F^{\mu\nu} ~+~ \pr_{\phi} \beta ~F_{\mu\nu}
~^\star F^{\mu\nu}   \label{gen-feq-b} \\
&&\nabla_{\mu} \left[\omega (\phi) \pr^{\mu} \xi\right] ~=~ \pr_{\xi} \alpha ~F_{\mu\nu}
F^{\mu\nu} ~+~ \pr_{\xi} \beta ~F_{\mu\nu} ~^\star F^{\mu\nu}   \label{gen-feq-c} \\
&&\nabla_{\mu} \left[\alpha (\phi, \xi) F^{\mu\nu} ~+~ \beta (\phi, \xi)
~^\star F^{\mu\nu}\right] ~=~ 0    \label{gen-feq-d} \\
&&\nabla_{\mu} ~^\star F^{\mu\nu} ~=~ 0.     \label{gen-feq-e}
\eea
\ese
The last equation in this set is the Maxwell-Bianchi identity.

We study the regular non-rotating charged black hole solutions of these field equations subject
to different limiting conditions. Our aim is to pinpoint the black hole families corresponding
to typical forms of the couplings $\omega$, $\alpha$ and $\beta$ in terms of the scalars $\phi$
and $\xi$. In the specific case of low energy effective string theory in four
dimensions, $\phi$ can be identified with the massless scalar dilaton while $\xi$ to 
the pseudoscalar axion. The low energy effective string action in the Einstein frame
corresponds to the above action (\ref{gen-action}) with $\omega = e^{2 \phi}, \alpha = 
e^{- \phi}$ and $ \beta = \xi$.

We look for static spherical symmetric black hole solutions with the line element of the form
\be    \label{line-element}
ds^2 ~=~ - N^2 (r) dt^2 ~+~ \frac{dr^2}{N^2 (r)} ~+~ R^2 (r) d \Omega^2 .
\ee
where ~$d \Omega^2 = (d \vth^2 + \sin^2 \vth d \vph^2)$~ is the metric on the two sphere.
The scalar fields $\phi$ and $\xi$ are functions of the radial variable $r$ only, while the
Maxwell's equations (\ref{gen-feq}d) and (\ref{gen-feq}e) can be integrated to yield the 
non-vanishing components
\bea \label{em-comp}
F_{\mu\nu}: && F_{tr} ~=~ - F_{rt} ~=~ \frac{q_e ~-~ q_m \beta}{\alpha R^2} ~dt \wedge dr\nno\\
&& F_{\vth\vph} ~=~ - F_{\vph\vth} ~=~ q_m \sin \vth ~d\vth \wedge d\vph
\eea
The integration constants $q_e$ and $q_m$ can respectively be interpreted as 
the scalar field shielded electric and magnetic charges (as of the isolated charges 
corresponding to a system with a scalar ``fluid'') defined in the usual way (see \cite{wald}):
\bea \label{shield-charge}
q_e ~=~ - \frac 1 {4 \pi} \int_S \left(\alpha~ ^\star F ~+~ \beta F\right)~;~~~~~~
q_m ~=~ - \frac 1 {4 \pi} \int_S ~F 
\eea
where the integrals are over a two-dimensional surface $S$, bounding a three-dimensional
hypersurface $\Sigma$, pushed to spatial infinity. A combination of $q_e$ and $q_m$ may,
in fact, be identified with the total (bare) electric and magnetic charges of the matter 
distribution once the asymptotic behaviour of the metric functions $R(r), N(r)$ and that 
of the couplings $\alpha(r), \beta(r)$ are ascertained. In asymptotically flat spacetimes, 
as will be shown later, such identifications are straightforward. However, in asymptotically 
non-flat spacetimes  a complete knowledge of the solutions to the field equations is required 
in order to identify $q_e$ and $q_m$ as combinations of the (bare) electric and magnetic 
charges. We shall discuss this in detail in the following sections.

In the static spherical symmetric spacetime the field equations (\ref{gen-feq}) reduce to
\bse
\label{full-feq}
\bea
&&(R^2 N^2)'' ~=~ 2 \label{full-feq-a}\\
&&\left(\frac{R'} R\right)' ~+ \left(\frac{R'} R\right)^2 ~+~ \frac 1 4
\left(\phi'^2 ~+~ \omega \xi'^2\right) =~ 0 \label{full-feq-b}\\
&&(R^2 N N')' ~=~ \frac 1 {R^2} \left[q_m^2 \alpha ~+~ \frac{(q_e ~-~ q_m \beta)^2}{\alpha}
\right] \label{full-feq-c}\\
&&(R^2 N^2 \omega \xi')' ~=~ \frac 2 {R^2} ~\pr_{\xi} \left[q_m^2 \alpha ~+~ \frac{(q_e ~-~
q_m \beta)^2}{\alpha}\right] \label{full-feq-d}\\
&&(R^2 N^2 \phi')' ~=~ \frac 2 {R^2} ~\pr_{\phi} \left[q_m^2 \alpha ~+~ \frac{(q_e ~-~
q_m \beta)^2}{\alpha}\right] ~+~ \frac 1 2 R^2 N^2 \xi'^2 \pr_{\phi} \omega  \label{full-feq-e}
\eea
\ese
where a prime denotes derivative with respect to $r$. Eq.(\ref{full-feq}a) can at once be
integrated to find that the product $R^2 (r) N^2 (r)$ is purely quadratic in $r$ such that,
\be \label{quad}
R^2 (r) N^2 (r) ~=~ (r ~-~ r_+)(r ~-~ r_-)
\ee
where $r_{\pm}$ are the  constants of integration. Eq.(\ref{full-feq}b), on the other hand, is
nothing but the trace equation corresponding to the Einstein's equations 
(\ref{gen-feq}a), which is essentially the scalar curvature equation. One may verify explicitly
that the possibility  of obtaining a black hole solution of the 
field equations in a closed form depends entirely on the integrability of this equation (\ref{full-feq}b).

Now, it is easy to check that entire set of field equations (\ref{full-feq}) can be
derived exactly from an effective action for Einstein-Maxwell theory coupled to a
single scalar field $\psi$ (which is defined in terms of $\phi$ and $\xi$ as $\psi'^2 
= \phi'^2 + \omega \xi'^2$):
\be \label{eff-action}
S_{eff} ~=~ \frac 1 {2 \kappa} \int d^4 x \sqrt{-g} \left[R ~-~ \frac 1 2 \pr_{\mu} \psi \pr^{\mu} 
\psi ~-~ \alpha (\psi) F_{\mu\nu} F^{\mu\nu} ~-~ \beta (\psi) F_{\mu\nu} ~^\star F^{\mu\nu}\right]
\ee
whence the effective field equations are given by
\bse
\label{eff-feq}
\bea
&&(R^2 N^2)'' ~=~ 2 \label{eff-feq-a}\\
&&\left(\frac{R'} R\right)' ~+ \left(\frac{R'} R\right)^2 ~+~ \frac 1 4 \psi'^2 ~=~ 0 \label{eff-feq-b}\\
&&(R^2 N N')' ~=~ \frac 1 {4 R^2} ~\gamma \label{eff-feq-c}\\
&&\left(R^2 N^2 \psi'\right)' ~=~ \frac 1 {2 R^2} ~\pr_{\psi} \gamma  \label{eff-feq-d}
\eea
\ese
where
\bse
\label{consis}
\bea
&&\gamma ~=~ 4\left[q_m^2 \alpha ~+~ \frac{(q_e ~-~ q_m \beta)^2}{\alpha}
\right] \label{consis-a}\\
&&\psi'^2 ~=~ \phi'^2 ~+~ \omega (\phi) \xi'^2 \label{consis-b}\\
&&\pr_{\xi} \gamma ~=~ 2 R^2 (R^2 N^2 \omega \xi')'.  \label{consis-c}
\eea
\ese

The equation (\ref{eff-feq}b) here is the scalar curvature equation which corresponds
to the trace of the Einstein's equations.
Charge neutrality of $\psi$ suggests the following ansatz for the metric  
\be \label{ansatz}
R (r) ~=~ r \left(k ~+~ \frac {2 r_0} r\right)^n
\ee
where $k$ and $r_0$ are constant real parameters; 
$k$ is dimensionless while $r_0$ has the dimension
of $r$. From Eq.(\ref{quad}) one therefore has
\be \label{N}
N (r) ~=~ \frac{\sqrt{\left(r ~-~ r_+\right) \left(r ~-~ r_-\right)}}{r \left(k ~+~ 2 r_0/r\right)^n}  .
\ee

It can be shown that Eq.(\ref{eff-feq}b) integrates to the following real
solution for $\psi(r)$, provided $n$ lies between $0$ and $1$: 
\be  \label{psi}
\psi (r) ~=~ \psi_0 ~-~ 2 \sqrt{n (1 - n)} ~ \ln \left(k ~+~ \frac {2 r_0} r\right) ~;~~~~~~~~~~
( 0 \leq n \leq 1 )
\ee
where $\psi_0$ is a constant of integration. For the extreme cases $n = 0$ or $1$, the field
$\psi$ is trivial and by suitably shifting the radial variable $r$ by a constant parameter 
given in terms of $k$ and $r_0$ one gets back the usual Reissner-Nordstr\"om (RN) solution of
Einstein-Maxwell theory. In order to construct a solution with a non-trivial $\psi$,
we therefore exclude these values and assume $0 < n < 1$.

The  equations (\ref{eff-feq}c) and  (\ref{eff-feq}d) together
yield the differential equation for the effective coupling parameter $\gamma$
\be \label{diffcoup}
\pr_{\psi}^2 \gamma ~-~ \frac {1 - 2 n}{\sqrt{n (1 - n)}} \pr_{\psi} \gamma ~-~ \gamma ~=~ 0
\ee
with solution
\be \label{coup1}
\gamma (r) ~=~ k_1 ~\exp \left[{\sqrt{\frac{1 - n} n} ~\psi (r)}\right] +~ k_2~ \exp \left[{- \sqrt{\frac n
{1 - n}} ~\psi (r)}\right]
\ee
where $k_1$ and $k_2$ are arbitrary constants. Using the above equation (\ref{psi}) for $\psi (r)$,
we have
\be \label{coup}
\gamma (r) ~=~ K_1 \left(k ~+~ \frac {2 r_0} r\right)^{2 (n - 1)} +~  K_2 \left(k ~+~ 
\frac {2 r_0} r\right)^{2 n}
\ee
where $K_1 = k_1 e^{\sqrt{(1 - n)/n}~ \psi_0}$ and $K_2 = k_2 e^{- \sqrt{n/(1 - n)} ~\psi_0}$.
Substituting this expression for $\gamma$ back in Eqs.(\ref{eff-feq}c) and  (\ref{eff-feq}d) and
solving for $K_1$ and $K_2$, we obtain
\be \label{K1K2}
K_1 ~=~ 4 n \left[4 r_0^2 ~+~ 2 k r_0 \left(r_+ ~+~ r_-\right) +~ k^2 r_+ r_-\right] ~;~~~~~
K_2 ~=~ 4 \left(1 - n\right) r_+ r_-  .
\ee

Now, one can verify that $k\neq 0~(k=0)$ in Eq.(\ref{ansatz}) 
corresponds to asymptotically flat (non-flat) spacetimes. For
the AF case, one can always set $k=1$, without loss of 
generality, to get black holes with spherical topology. 
In what follows, we determine the masses for both AF and AN solutions
corresponding to the cases $k = 1$ and $k = 0$ respectively.

\section{The mass formulae \label{massform}}

We resort to the general formalism of quasi-local mass developed 
independently by Brown
and York (BY) \cite{brown} and by Hawking and Horowitz (HH) \cite{hawkhor}, 
and further described
in \cite{mann,mann1,chan} in the context of charged black holes 
with curved asymptotes
in (2+1) dimensional \cite{mann1} 
and (3+1) dimensional dilaton gravity \cite{mann}. 
Conformal
invariance of the quasi-local mass, 
according to both BY and HH prescriptions, has been studied
in detail in \cite{chan} and alternative proposals have been put forward 
in \cite{sbose}.
General features of quasi-local mass of non-asymptotically flat 
dilaton black holes have also
been discussed recently in \cite{clement}. 
We briefly outline the following rule:

\bigskip
In $D$ dimensions, given a generic action functional of the 
metric $g_{\mu\nu}^D$ and scalar and other matter fields $\{\psi, F\}$:
\be \label{act-gen}
S (g_D, \psi, F) ~=~ \frac 1 {2 \kappa_D} \int_{\cal M} d^D x \sqrt{- g_D} ~U (\psi)
\left[{\cal R} (g_D) - W (\psi) \pr_\mu \psi \pr^\mu \psi - V (\psi) +
X (\psi) L_m (F)\right]
\ee
and a line element of the form
\be \label{met-gen}
ds^2 ~=~ - N^2 (R) dt^2 ~+~ \frac{dR^2}{Y^2 (R)} ~+~ R^2 d \Omega_{D-2}^2
\ee
the quasi-local mass is defined as
\be \label{quasi-gen}
M (R_b) ~=~ \frac{N (R_b)}{\kappa_D} \left[Y_0 (R_b) ~-~ Y (R_b)\right] \left[ \frac d {dR}
\left\{A_{D-2} (R) ~U (\psi)\right\} \right]_{R = R_b}
\ee
where $\kappa_D = 8 \pi G_D$ is the $D$ dimensional gravitational coupling constant;
$Y_0 (R_b)$ is an arbitrary function which determines the zero of energy of a
background spacetime; and $R_b$ is the radius of the $(D-2)$ dimensional spacelike
hypersurface boundary whose area is $A_{D-2}$ --- for a boundary with the topology
of a sphere
\be
A_{D-2} ~=~ (4 \pi)^{\frac{D-2} 2} ~\frac{\Gamma \left({\frac{D-2} 2}\right)}
{\Gamma \left(D - 2\right)} ~ R_b^{D-2} ~=~ \int_{\cal B} d^{D-2} x \sqrt{\sigma}
\ee
where $\sigma$ is the determinant of the metric on the boundary ${\cal B}$.

In four dimensions ($D = 4$), the area of a spherical hypersurface boundary is simply given
by $A_{D-2} \equiv A_2 = 4 \pi R_b^2$, and the expression (\ref{quasi-gen}) for the quasi-local
mass reduces to
\be \label{quasi4}
m (R_b) ~\equiv~ G M (R_b) ~=~ N (R_b) \left[Y_0 (R_b) ~-~ Y (R_b)\right]
\left[\frac{R^2} 2 \left(\frac{d U (\psi)}{dR} ~+~ \frac{2 U (\psi)} R\right)\right]_{R = R_b} .
\ee

Now, comparing the general action (\ref{act-gen}) with the effective action (\ref{eff-action})
for $D = 4$, we have $U (\psi) = 1$ and $W (\psi) = 1/2$, whence the quasi-local mass is given
by
\be \label{quasi}
m (R_b) ~=~ R_b N (R_b) \left[Y_0 (R_b) ~-~ Y (R_b)\right] .
\ee
Also, casting the metric (\ref{met-gen}) in the form (\ref{line-element}), one has
$Y = N dR/dr$. Setting $r = r_b$ corresponding to $R = R_b$, we have
\be \label{quasifinal}
m (r_b) ~=~ R (r_b) N (r_b) \left[Y_0 (r_b) ~-~ Y (r_b)\right] .
\ee
If $R$ is a monotonically increasing function of $r$, and vice versa, the actual gravitating
rest mass $m ( \equiv G M )$ may be obtained by taking the value of $m (r_b)$ in the limit
$r_b \rightarrow \infty$, i.e., by pushing the hypersurface boundary to spatial infinity.
Indeed in the present scenario, the function $R$, given by the ansatz (\ref{ansatz}),
increases monotonically with $r$ for all values of the parameters $k$ and $r_0$, since
as mentioned above the index $n$ can only take values strictly between $0$ and $1$ for
a real solution of the effective field $\psi$. Using Eq.(\ref{N}) for the metric function 
$N (r)$, we get
\be \label{Y}
Y (r) ~=~ N (r) \frac d {dr} R (r) ~=~ \left(1 ~-~ \frac{2 n r_0}{k r + 2 r_0}\right) 
\sqrt{\left(1 ~-~ \frac{r_+} r\right) \left(1 ~-~ \frac{r_-} r\right)} .
\ee
As such, the expression for the function $Y_0 (r_b)$, and hence the quasi-local mass
$m (r_b)$, can be obtained in the AF ($k = 1$) and AN ($k = 0$) cases as follows:

\bigskip
The general expression for the background (or reference) line-element is given by
\be   \label{ref-metric}
ds_0^2 ~=~ - N_0^2 dt^2 ~+~ \frac{dr^2}{Y_0^2} ~+~ r^2 d \Omega^2 .
\ee

\bigskip
\noindent
{\bf (i)}~~ For the AF case ($k = 1$) the background may universally be chosen \cite{chan,sbose}
to be flat with a constant lapse
\be
N_0 ~\equiv~ N (r_b) ~=~ \frac{\sqrt{\left(r_b ~-~ r_+\right) \left(r_b ~-~ r_-\right)}}
{r_b \left(1 ~+~ 2 r_0/r_b\right)^n}
\ee
as dictated by the HH prescription \cite{hawkhor}; and $Y_0 = 1$. This background metric is in
fact the solution of the field equations corresponding to the action (\ref{eff-action}) for suitable
reference values of the scalar $\psi$ and the gauge field $F$, viz., $\psi = \psi_0$ and $F = 0$
respectively. The quasi-local mass can be evaluated using Eqs.(\ref{quasifinal}), (\ref{Y}) and
(\ref{N}) as
\be \label{quasi1}
m (r_b) ~=~ r_b \left[\sqrt{\left(1 ~-~ \frac{r_+}{r_b}\right) \left(1 ~-~ \frac{r_-}{r_b}\right)}
~-~ \left(1 ~-~ \frac{2 n r_0}{r_b + 2 r_0}\right) \left(1 ~-~ \frac{r_+}{r_b}\right)
\left(1 ~-~ \frac{r_-}{r_b}\right)\right].
\ee
Taking the limit $r_b \rightarrow \infty$, we get
\be \label{mass1}
m ~=~ \lim_{r_b \rightarrow \infty} m (r_b) ~=~ \frac{r_+ ~+~ r_-} 2 ~+~ 2 n r_0
\ee
which is the same as the gravitating mass one could have obtained in the Arnowitt-Deser-Misner (ADM)
formalism in asymptotically flat spacetime.

\bigskip
\noindent
{\bf (ii)}~~ For the AN case ($k = 0$) there is no unique choice of the reference geometry.
However, a simple choice \cite{chan,sbose,mann} for the background spacetime 
is the line element obtained by setting the constants $r_\pm$ to zero 
in the (non-flat) space part of the original line element (\ref{line-element}). 
In other words, the background metric is chosen to be that given by Eq.(\ref{ref-metric}) with
\bea
N_0 &\equiv& N (r_b) ~=~ (2 r_0)^{- n} r_b^{n - 1} \sqrt{(r_b ~-~ r_+)(r_b ~-~ r_-)}
~~~~~~~~~~~\hbox{and} \nno\\
Y_0 &\equiv& Y (r_b) \vert_{r_\pm = 0} ~=~ 1 - n  ~~~~~~~~~~~~~~~~~~~~~~~~~
~~~~\hbox{[from Eq.(\ref{Y})]} .
\eea
The quasi-local mass turns out to be
\be \label{quasi0}
m (r_b) ~=~ (1 - n)~ r_b \left[\sqrt{\left(1 ~-~ \frac{r_+}{r_b}\right)
\left(1 ~-~ \frac{r_-}{r_b}\right)} ~-~ \left(1 ~-~ \frac{r_+}{r_b}\right)
\left(1 ~-~ \frac{r_-}{r_b}\right)\right]
\ee
whence taking the limit $r_b \rightarrow \infty$ one obtains the gravitating rest mass
\be \label{mass0}
m ~=~ \lim_{r_b \rightarrow \infty} m (r_b) ~=~ \left(\frac{1 - n} 2\right) \left(r_+ ~+~
r_-\right) .
\ee

\section{General structures of static spherically symmetric black hole solutions
with or without flat asymptotes    \label{gensol}}

In this section we outline the general forms of both AF and AN solutions
representing static charged non-rotating black holes in scalar coupled
Einstein-Maxwell theory in four dimensions.

\subsection{Asymptotically flat solutions  \label{afsol}}

Let us recall the whole set [(\ref{eff-feq}) -- (\ref{K1K2})] of effective field equations
and the solutions that have been obtained with the metric ansatz (\ref{ansatz}) in section
\ref{general}. We set the parameter $k = 1$ and refer to the mass formula (\ref{mass1}) 
given in section \ref{massform} for AF spacetimes. Solving the equations (\ref{K1K2}) 
one obtains the expressions for
$r_\pm$:
\be  \label{1-rpma}
r_\pm ~=~ - r_0 ~+~ \frac 1 {16 r_0} \left(\frac{K_1} n - \frac{K_2}{1 - n}\right) \pm~
\sqrt{r_0^2 + \frac 1 {256 r_0^2} \left(\frac{K_1} n - \frac{K_2}{1 - n}\right)^2 -
\frac 1 8 \left(\frac{K_1} n + \frac{K_2}{1 - n}\right)}
\ee
whence the mass expression (\ref{mass1}) reduces to
\be
m ~=~ \frac 1 {16 r_0} \left(\frac{K_1} n ~-~ \frac{K_2}{1 - n}\right) +
\left(2 n ~-~ 1\right) r_0.
\ee

Under a coordinate shift $r + r_0 \rightarrow r$ and a redefinition
$r_\pm + r_0 \Rightarrow r_\pm$, the line element retains its original form (\ref{line-element}),
however, with metric functions $N (r)$ and $R (r)$ given as
\bea \label{1-metric}
N (r) ~=~ \frac{\sqrt{(r - r_+) (r - r_-)}}{R (r)} ~,~~~~~ R (r) ~=~ (r - r_0)^{1 - n}
(r + r_0)^n
\eea
where
\bea \label{1-rpm}
r_\pm &=& m_0 ~\pm~ \sqrt{m_0^2 ~+~ r_0^2 ~-~ \frac 1 8 \left(\frac{K_1} n ~+~
\frac{K_2}{1 - n}\right)}
\eea
with
\bea \label{1-r0m0}
r_0 &=& \frac 1 {16 m_0} \left(\frac{K_1} n ~-~ \frac{K_2}{1 - n}\right) ~;~~~~
m_0 ~=~ m ~-~ (2 n - 1) ~r_0 .
\eea
Asymptotic expansion of (\ref{1-metric}), using (\ref{1-rpm}) and (\ref{1-r0m0}) suggests 
that $m$ is the ADM mass of the charged matter distribution. The solutions (\ref{psi}) 
for the effective scalar $\psi$ and (\ref{coup}) for the effective coupling $\gamma$ 
reduce respectively to
\bse
\label{1-psicoup}
\bea
&&\psi (r) ~=~ \psi_0 ~+~ 2 \sqrt{n \left(1 - n\right)} ~\ln \left(\frac{r ~-~ r_0}
{r ~+~ r_0}\right) \label{1-psicoup-a}\\
&&\gamma (r) ~=~ K_1 \left(\frac{r ~-~ r_0}{r ~+~ r_0}\right)^{2 (1 - n)} +~
K_2 \left(\frac{r ~-~ r_0}{r ~+~ r_0}\right)^{- 2 n}.  \label{1-psicoup-b}
\eea
\ese

The Ricci, squared-Ricci and Kretschmann scalar invariants all fall off to zero as
$r \rightarrow \infty$:
$$
{\cal R} ~\sim~ \frac 1 {r^4} ~;~~~~~~~~~~ {\cal R}_{\mu\nu} {\cal R}^{\mu\nu} ~\sim~ \frac 1 {r^8}~;
~~~~~~~~~~ {\cal K} ~\equiv~ {\cal R}_{\mu\nu\lambda\sigma} {\cal R}^{\mu\nu\lambda\sigma} ~\sim~
\frac 1 {r^6} .
$$
In fact for large $r$, the static spherical symmetric metric
given in Eq.(\ref{1-metric}) can always be cast
in the form $g_{\mu\nu} = \eta_{\mu\nu} + {\cal O} (1/r)$ along either
spatial or null directions, $\eta_{\mu\nu}$ being the Minkowskian.
Also, equations (\ref{1-psicoup}) show that the effective scalar $\psi$
as well as the effective coupling $\gamma$, tend toward constant
values ($\psi_0$ and $K_1 + K_2$ respectively) as $r \rightarrow \infty$,
thus verifying that the spacetime is asymptotically flat.

Now, the conditions for existence of regular black hole event horizon(s)
can be given as follows:

\begin{itemize}
\item Firstly, the positivity of the mass $m$ imposes a restriction the value
of the parameter $m_0$:
\bea \label{1-cond1}
m_0 ~>~ \left(1 ~-~ 2 m\right) r_0 .
\eea
\item Secondly, the term under the square root in the expression (\ref{1-rpm})
must be non-negative. Using Eqs.(\ref{1-r0m0}), this implies
\bea \label{1-cond2}
\left(m_0 ~-~ r_0\right)^2 ~\geq~ \frac{K_2}{4 \left(1 - n\right)} .
\eea
\item Finally, in order that at least one horizon exists, one must have $r_+ > r_0$,
otherwise the metric function $N (r)$, given by Eq.(\ref{1-metric}), will blow up
before a radially infalling particle reaches the surface $r = r_+$ from infinity. Indeed,
$r = r_0$ is a curvature singularity since all the scalar invariants ${\cal R},
{\cal R}_{\mu\nu} {\cal R}^{\mu\nu}$ and ${\cal R}_{\mu\nu \alpha\beta}
{\cal R}^{\mu\nu\alpha\beta}$ diverge therein, whereas the metric function $R (r = r_0)$
vanishes, as $0 < n < 1$ [See Eqs.(\ref{1-metric})]. Now, $R (r)$ is the radius of the
two-spherical hypersurface of the spacetime given by the line element (\ref{line-element}).
Therefore, as $R$ vanishes at $r = r_0$, the two-sphere crushes to a point, thereby
implying that the singularity at $r = r_0$ is a `point' (curvature) singularity. Such a
singularity must obviously be naked unless at least there is a horizon at $r_+ (> r_0)$.
From Eqs.(\ref{1-rpm}) and (\ref{1-r0m0}) this means
\bea \label{1-cond3}
\left(m_0 ~-~ r_0\right) +~ \sqrt{\left(m_0 ~-~ r_0\right)^2 -~ \frac{K_2}{4 \left(1 - n\right)}} ~>~ 0 .
\eea
\end{itemize}

Let us consider separately the following situations:

\bigskip
\noindent
{\bf (i) $m_0 > r_0$ :} ~ The condition (\ref{1-cond1}) is satisfied automatically for any $n$ greater than
zero, which is always true of course. The condition (\ref{1-cond2}) implies
$$ m_0 ~\geq~ r_0 ~\pm~ \frac 1 2 ~\sqrt{\frac{|K_2|}{1 - n}} .$$
The inequality for the lower ($-$) sign is always valid since we are considering the case $m_0 > r_0$.
However, the upper ($+$) sign sets a lower bound (LB) on $m_0$:
\be
m_0^{LB} ~=~ r_0 ~+~ \frac 1 2 ~\sqrt{\frac{|K_2|}{1 - n}}
\ee
which, in turn, implies a lower bound on the mass $m$:
\bea \label{1-mlb}
m^{LB} ~=~ 2 n r_0 ~+~ \frac 1 2 ~\sqrt{\frac{|K_2|}{1 - n}} .
\eea
In any case, regardless of this bound, the condition (\ref{1-cond3}) is always fulfilled whenever
$m_0 > r_0$. However, the above bound on the mass definitely ensures that a blackhole
horizon exists for $m \geq m^{LB}$.

\bigskip
\noindent
{\bf (ii) $m_0 = r_0$ :} ~ The condition (\ref{1-cond1}) implies $n > 0$, which is always valid.
However, since $n < 1$, both the conditions (\ref{1-cond2}) and (\ref{1-cond3}) are satisfied,
and as such an event horizon exists, only for $K_2 < 0$.

\bigskip
\noindent
{\bf (ii) $m_0 < r_0$ :} ~ The condition (\ref{1-cond1}) again implies $n > 0$, which is true. The
condition (\ref{1-cond2}) is satisfied for
$$ r_0 ~\geq~ m_0 ~\pm~ \frac 1 2 ~\sqrt{\frac{|K_2|}{1 - n}} .$$
The inequality for the lower ($-$) sign is always valid for any $m_0$ less than $r_0$; however,
the upper ($+$) sign sets an upper bound (UB) on $m_0$:
\be
m_0^{UB} ~=~ r_0 ~-~ \frac 1 2 ~\sqrt{\frac{|K_2|}{1 - n}}
\ee
that is, an upper bound on the mass $m$:
\bea \label{1-mub}
m^{UB} ~=~ 2 n r_0 ~-~ \frac 1 2 ~\sqrt{\frac{|K_2|}{1 - n}} .
\eea
The condition (\ref{1-cond3}) further imposes the restriction $K_2 < 0$. Thus on the whole, the
existence of a blackhole horizon for this case is ensured definitely when $m \leq m^{UB}$ and
$K_2 < 0$.

\bigskip

In section \ref{dilax}, in order to find the complete solution,
we find expressions (as functions of $r$) for our original set of
scalar fields $\{ \phi, \xi \}$, from the set of solutions (\ref{1-metric})
-- (\ref{1-psicoup}). We also relate the constants $K_1, K_2$ with the
electromagnetic charges $q_e, q_m$ and the asymptotic values ($\phi_0$ and
$\xi_0$, say) of the scalars $\phi$ and $\xi$. To this end, we check the
compatibility of the entire set (\ref{1-metric}) -- (\ref{1-psicoup}) with 
the set of equations (\ref{consis}) and solve the latter for $\phi$ and $\xi$.

However, the issue is easily settled in a specific situation where the
functional forms of the couplings $\alpha$ and $\beta$ (and as such $\gamma$) 
are known in terms of $\phi$ and $\xi$. We discuss this in detail in section 
\ref{dilax} as we analyze the solutions for various cases corresponding to 
the specific scenario of dilaton-axion coupled Einstein-Maxwell theory.

\subsection{Asymptotically non-flat solutions  \label{acsol}}

Let us again refer to the set [(\ref{eff-feq}) -- (\ref{K1K2})] of effective field equations
and solutions obtained in section \ref{general}. We consider the case $k = 0$
in the solution ansatz (\ref{ansatz}), and resort to the corresponding expression (\ref{mass0})
for the gravitating mass derived in section \ref{massform}. The metric functions $N (r)$ and
$R (r)$ are given by
\bea \label{2-metric}
N (r) ~=~ \frac{\sqrt{(r - r_+) (r - r_-)}}{R (r)} ~,~~~~~~~~~~ R (r) ~=~ 
r \left(\frac {2 r_0} r\right)^n .
\eea
Solving simultaneously the equations (\ref{K1K2}) ~(for $k = 0$)~ and  (\ref{mass0}), we obtain
\be  \label{2-rpm}
r_\pm ~=~ \left(\frac 1 {1 - n}\right) \left[m ~\pm~ \sqrt{m^2 ~-~ (1 - n) \frac{K_2} 4}\right] .
\ee

Since the parameter $n$ lies between $0$ and $1$, one can check explicitly the large $r$ fall-offs
of all the scalar invariants
$$
{\cal R} ~\sim~ r^{2 (n - 1)} ~;~~~~~~~~~~ {\cal R}_{\mu\nu} {\cal R}^{\mu\nu} ~\sim~ r^{4 (n - 1)}~;
~~~~~~~~~~ {\cal K} ~\equiv~ {\cal R}_{\mu\nu\lambda\sigma} {\cal R}^{\mu\nu\lambda\sigma} ~\sim~
r^{4 (n - 1)} .
$$
However, the spacetime is not asymptotically flat since the metric component
$|g_{tt}| = |g^{rr}| = N^2 (r)$ goes as $r^{2 n}$ in the limit $r \rightarrow \infty$.
The solutions (\ref{psi}), for the effective scalar $\psi$,  and (\ref{coup}), for the
effective coupling $\gamma$, which now reduce to
\bse
\label{2-psicoup}
\bea
&&\psi (r) ~=~ \psi_0 ~-~ 2 \sqrt{n \left(1 - n\right)} ~\ln \left(\frac {2 r_0} r\right) \label{2-psicoup-a}\\
&&\gamma (r) ~=~ \left(4 n r^2 ~+~ K_2\right) \left(\frac {2 r_0} r\right)^{2 n} ,  \label{2-psicoup-b}
\eea
\ese
also diverge as  $r \rightarrow \infty$. Note that in deriving Eq.(\ref{2-psicoup}b) we have
substituted $16 n r_0^2$ for $K_1$, following Eq.(\ref{K1K2}).

The possible lower bound on the mass $m$ for the existence of event horizon(s) can be obtained
from Eqs.(\ref{2-rpm}):
\be
m^{LB} ~=~ \frac 1 2 ~\sqrt{\left(1 ~-~ n\right) |K_2|} .
\ee
Moreover, the divergent behaviour of the effective scalar $\psi$  and the effective coupling
$\gamma$ as $r \rightarrow \infty$, is reflected in similar divergences of the original set of
scalar fields $\{ \phi, \xi \}$ in the asympototic limit, which we shall show later on in the context
of dilaton-axion blackholes.

In order to determine the precise functional forms of the scalar fields $\{ \phi, \xi \}$ and
to ascertain the value of the arbitrary constant $K_2$, we again need to check the
compatibility of the entire set of solutions (\ref{2-metric}) -- (\ref{2-psicoup}) with the
equations (\ref{consis}) and solve the latter in some specific circumstances where the
functional forms of the couplings $\alpha$ and $\beta$ (and as such $\gamma$) are known
in terms of $\phi$ and $\xi$. In the following section we resort to the dilaton-axion coupled
Einstein-Maxwell theory, from the perspective of low energy effective string theory (with or
without certain parametric modifications/generalizations), and study the charged black hole
solutions corresponding to some typical special cases.

\section{Charged black holes in Einstein-Maxwell-dilaton-axion theory
\label{dilax}}

Let us refer to a generalized action for Einstein-Maxwell
theory in four dimensions, coupled to the
massless scalar dilaton $\phi$ and the massless pseudoscalar
axion $\xi$ in Einstein frame:
\be \label{action-dilax}
S ~= \int d^4 x \sqrt{-g} \left[\frac 1 {2 \kappa} \left(R - \frac 1 2 \pr_{\mu} \phi \pr^{\mu} \phi
- \frac 1 2 e^{2 a \phi} \pr_{\mu} \xi \pr^{\mu} \xi\right) - e^{- a \phi}
F_{\mu\nu} F^{\mu\nu} -  b \xi F_{\mu\nu} ~^\star F^{\mu\nu}\right].
\ee
where $a$ and $b$ are two constant free parameters, the presence of which
generalizes the low energy effective string action
compactified to four spacetime dimensions.
Both the dilaton and the axion fields are kept dimensionless.

For the ten dimensional low energy heterotic superstring compactified 
on a six-torus, the bosonic
part of the effective action corresponds to (\ref{action-dilax}) with the parameters $a = b = 1$.
The axion $\xi$ is connected via a duality in four dimensions with the three form $H_{\mu\nu\lambda}$,
which is the field strength corresponding to the Kalb-Ramond (KR) two form $B_{\mu\nu}$, appearing
in the string massless spectrum, augmented with the Chern-Simons term $A_{[\mu} F_{\nu\lambda]}$ that
is required to keep the underlying theory free from gauge anomaly:
\be \label{axion}
H_{\mu\nu\lambda} ~=~ \pr_{[\mu} B_{\nu\lambda]} ~+~ 2 \sqrt{2 \kappa}~ A_{[\mu} F_{\nu\lambda]}
~=~ \frac{e^{2 a \phi}}{2 \kappa} ~\epsilon_{\mu\nu\lambda}^{~~~~ \sigma} ~\pr_\sigma \xi .
\ee
The four dimensional heterotic string action is expressed in terms of the dilaton $\phi$
and the three form $H_{\mu\nu\lambda}$ as
\be \label{het-action}
S_H ~=~ \int d^4 x \sqrt{-g} \left[\frac 1 {2 \kappa}\left(R ~-~ \frac 1 2 \pr_{\mu} \phi \pr^{\mu}
\phi\right) -~ \frac 1 {12} e^{- 2 \phi} H_{\mu\nu\lambda} H^{\mu\nu\lambda} ~-~ e^{- \phi}
F_{\mu\nu} F^{\mu\nu}\right]
\ee
the effective version of which in terms of $\phi$ and $\xi$ is that given in Eq.(\ref{action-dilax})
with the parameters $a$ and $b$ set to unity. In our subsequent analysis, we shall be referring
both to this conventional string theoretic picture, and the generalized scenarios where either
$a$ or $b$ or both differ(s) from unity.

Deviations from usual the four dimensional low energy effective string action (\ref{het-action})
primarily arise through the typical schemes of compactification (inspite of the standard
Kaluza-Klein dimensional reduction) of the ten dimensional superstring to four spacetime
dimensions. The invariance of the string effective equations of motion under certain symmetry
groups and the duality conjectures, viz., both T-duality and strong-weak coupling duality, have
been exploited in the context of different compactification schemes in order to develop the
generating technique of non-trivial solutions from known field configurations \cite{stw}-\cite{sen1}.
In fact, the dual rotations and constant shifts in the axion that generates the $SL (2,R)$ group
of symmetries of the equations of motion of dimensionally reduced heterotic string in the
perturbative sector, have been shown to inflict a mutual inter-rotation of electric and magnetic
charges as well as the dilaton and axion charges. This, in turn, enables one to obtain new
solutions with non-trivial axion and dilaton starting from known ones with only non-trivial dilaton
field but no axion \cite{stw,sen,ortin}. With the involvement of scalar and vector moduli fields, as
shown in \cite{sen,sen1}, there may still remain a symmetry of the equations of motion which
can be utilized in the solution generating process. In a plethora of works \cite{sen}-\cite{kalortin},
\cite{horstrom}-\cite{gl} such new solutions have been explored which in occasions describe
regular black holes and/or black strings with various characterizations --- charged, rotating or
non-rotating --- in presence of the dilaton and may also the axion (or, else the KR field) in different
spacetime dimensions. We, however, in this paper confine our study to some parametric
generalizations of the usual string picture, described by the action (\ref{het-action}), through the
introduction of the parameters $a$ and $b$. The motivations behind such generalizations are as follows:

\bigskip
The presence of the parameter $a$, which we assume to be a real and non-negative constant,
may be considered as a regulator for the strength of dilaton coupling with the Maxwell field.
Such a parameter was originally taken into account 
in the works of Gibbons and Maeda
\cite{gib,gibm} in the context of dilaton black holes and membranes 
in higher dimensions. A
number of interesting special cases may be pointed out:~
\begin{itemize}
\item The case $a = 1$, as mentioned above, 
corresponds to the field theoretic limit of the
ten dimensional or the effective four dimensional superstring model, 
in the bosonic sector.
\item The case $a = \sqrt{1 + 2/n}$ corresponds to the four 
dimensional Kaluza-Klien
toroidal reduction of a $4 + n$ dimensional theory.
\item The case $a = 0$, corresponds to the usual Einstein-Maxwell 
theory coupled minimally
with a massless Klien-Gordon scalar field $\phi$, 
whenever the presence of the other scalar
$\xi$ (or, else the KR three form $H_{\mu\nu\lambda}$) is ignored.
\end{itemize}

\bigskip
The appearance of $b$ may be related to the fact that the 
KR three tensor
$H_{\mu\nu\lambda}$ can be interpreted as the torsion of background 
spacetime.
Following the formalism in \cite{pmssg} one can identify
$H_{\mu\nu\lambda}$ with a completely antisymmetric 
torsion tensor $T_{\mu\nu\lambda}$,
modulo some multiplicative constant 
which regulates the strength of torsion-KR coupling.
The presence of the Chern-Simons term in the definition 
of $H_{\mu\nu\lambda}$ is 
crucial in restoring the $U(1)$ gauge invariance,
which is apparently lost in the coupling
of torsion with the $U(1)$ gauge field \cite{pmssg}. 
Symbolically, $T_{\mu\nu\lambda} = $
constant $ \times H_{\mu\nu\lambda}$, and the effective four dimensional version of
the theory in terms of the dual axion field $\xi$, can in general involve a constant parameter
in front of the kinetic term of $\xi$ as in Eq.(\ref{action-dilax}) (see \cite{ssgsscosm} for
a detailed analysis). Under suitable scaling of the $\xi$ field, one can get rid of the constant,
thereby restoring the canonical form of the kinetic term of $\xi$. In the process, however,
it only leads to another constant parameter $b$ being invoked in the $\xi F ~^\star F$ coupling
term in the string effective action as in (\ref{action-dilax}).

\bigskip
The introduction of both the parameters $a$ and $b$ may also be motivated from another interesting
viewpoint. For specific values of $a$ and $b$, the entire action (\ref{action-dilax}) is shown
\cite{dmssgssnew} to lead exactly to the field equations corresponding to a four dimensional
effective compactified version of a higher dimensional (bulk) Einstein-Maxwell-dilaton-Kalb-Ramond
theory in Randall-Sundrum (RS) scenario \cite{rs} (that provides a resolution to the well-known
Planck-electroweak hierarchy problem). The standard Kaluza-Klien decomposition of all the bulk
fields gives rise to both massless and towers of massive projections on the four dimensional
subspace --- the `3-brane' --- in which all the standard model fields are supposed to be confined.
Demanding that all the massive modes must lie in the TeV range in a RS picture, it has been found
that the massless modes of both the bulk KR field \cite{bmsenssgPRL,ssg,dmssg} and the scalar dilaton
\cite{dmssgssnew} are suppressed by a huge exponential (RS `warp') factor on the 3-brane, while
the massless mode of the gauge field suffers no such suppression \cite{dhr}. The net result
in a low energy effective four dimensional theory is that the exponential dilaton coupling with
the KR field and the gauge field is virtually washed out, while in contrast, the strength of the
KR-electromagnetic interaction, via the Chern-Simons extension, is immensely boosted by the
large warp factor. Expressing the KR massless mode in terms of the axion, one finds that the four
dimensional effective action corresponds exactly with (\ref{action-dilax}) for $a \sim e^{- \sigma} \ll 1$
and $b \sim e^{\sigma} \gg 1$ \cite{dmssgssnew}, where $\sigma$ is the RS warping.

\bigskip
Thus explaining the possible origin of the constant parameters $a$ and $b$ and their respective values
in different circumstances, let us now look for complete black hole descriptions of the solutions of field
equations in various situations in the context of dilaton-axion coupled Einstein-Maxwell theory.
For convenience, we choose to work in the natural units setting ~$2 \kappa = 16 \pi G = 1$.

A comparison of the action (\ref{action-dilax}) with the original action (\ref{gen-action}) reveals
the following coupling specifications:
\be \label{couplings}
\omega (\phi) ~=~ e^{2 a \phi}~;~~~~~~ \alpha (\phi) ~=~ e^{- a \phi}~;~~~~~~ \beta (\xi) ~=~ b ~\xi
\ee
and shows that  the set of equations (\ref{consis}), given in section \ref{general}, now take the form
\bse
\label{fconsis}
\bea
&&\gamma ~=~ 4 ~q_m^2 ~ e^{- a \phi} ~+~ 4 \left(q_e ~-~ q_m~ b~ \xi\right)^2
e^{a \phi} \label{fconsis-a}\\
&&\psi'^2 ~=~ \phi'^2 ~+~ e^{2 a \phi} \xi'^2 \label{fconsis-b}\\
&&4 ~ q_m b \left(q_e ~-~ q_m b~ \xi\right) e^{a \phi} =~ - R^2 \left[\left(r ~-~ r_+\right)
\left(r ~-~ r_-\right) e^{2 a \phi}~ \xi'\right]'.  \label{fconsis-c}
\eea
\ese
In what follows, we carefully examine the compatibility of
the general expressions for the metric functions given in sections
\ref{afsol} and \ref{acsol} for AF and AN spacetimes and the
corresponding expressions for the effective scalar and the effective
coupling ($\psi$ and $\gamma$ respectively), with the above set of
equations (\ref{fconsis}). Resorting separately to the AF and AN
cases, we try to determine the functional forms of the scalars $\phi$
and $\xi$  by solving Eqs.(\ref{fconsis}) in some specific situations
corresponding to some typical values of the parameters $a$ and $b$.

\subsection{Asymptotically flat black holes        \label{dilax-af}}

We list below the full set of solutions obtained in section \ref{afsol} for the metric,
the effective scalar $\psi$ and the effective coupling $\gamma$, as well as the
expressions for the electromagnetic field components, given in section \ref{general}.

\bigskip
\noindent
{\bf Line element:} \label{af-line}
\bea
ds^2 = - \frac{\left(r - r_+\right)\left(r - r_-\right)}{\left(r
- r_0\right)^{2 - 2 n} \left(r + r_0\right)^{2 n}} dt^2 +
\frac{\left(r - r_0\right)^{2 - 2 n} \left(r + r_0\right)^{2 n}}
{\left(r - r_+\right)\left(r - r_-\right)} dr^2 + \frac{\left(r
+ r_0\right)^{2 n}}{\left(r - r_0\right)^{2 n - 2}} d\Omega^2
\eea
{\bf Effective scalar:}
\bea \label{af-effscalar}
\psi (r) ~=~ \psi_0 ~+~ 2 \sqrt{n \left(1 - n\right)} ~\ln \left(\frac{r ~-~ r_0}
{r ~+~ r_0}\right)
\eea
{\bf Effective coupling:}
\bea \label{af-effcoup}
\gamma (r) ~=~  K_1 \left(\frac{r ~-~ r_0}{r ~+~ r_0}\right)^{2 (1 - n)} +~
K_2 \left(\frac{r ~-~ r_0}{r ~+~ r_0}\right)^{- 2 n}
\eea
{\bf Electromagnetic field components:}
\bea \label{af-em}
F_{\mu \nu} :~~~~ F_{tr} ~=~ \frac{\left(q_e ~-~ q_m b ~\xi\right) e^{a \phi}}
{\left(r - r_0\right)^{2 (1 - n)} \left(r + r_0\right)^{2 n}} ~dt \wedge dr~;
~~~~ F_{\vth\vph} ~=~ q_m \sin \vth ~d\vth \wedge d\vph
\eea
where
\bea \label{af-rpm}
r_\pm ~=~ m_0 ~\pm~ \sqrt{m_0^2 + r_0^2 - \frac 1 8 \left(\frac{K_1} n +
\frac{K_2}{1 - n}\right)} ~; \nno\\
r_0 ~=~ \frac 1 {16 m_0} \left(\frac{K_1} n - \frac{K_2}{1 - n}\right) ~;~~~~
m_0 ~=~ m - (2 n - 1) r_0 .
\eea
The relationships of $\{\psi, \gamma\}$ with $\{\phi, \xi\}$ are given by
Eqs.(\ref{fconsis}a) and (\ref{fconsis}b). Letting the scalar fields $\phi$
and $\xi$ approach the limiting values $\phi_0$ and $\xi_0$, as $r \rightarrow
\infty$,  in an asymptotically flat (AF) spacetime, it is possible to relate
the scalar-shielded electric and magnetic charges $q_e$ and $q_m$ to the
total (bare) electric and magnetic charges $Q_e$ and $Q_m$. These are defined
through the relations \cite{wald}
\be  \label{em-charge}
Q_e = - \frac 1 {4 \pi} \int_S ~^\star F = - \frac 1 {4 \pi} \int_S  E_r v^r dS~;~~~~~~~~
Q_m = - \frac 1 {4 \pi} \int_S ~ F = \frac 1 {4 \pi} \int_S  B_r v^r dS
\ee
where the two-sphere $S$ is the boundary of a three-dimensional hypersurface
$\Sigma$, pushed to spatial infinity; and the integrals are of the normal (radial)
components of the electric and magnetic vectors $E_\alpha = F_{\alpha\beta} n^\beta$
and $B_\alpha = - ~^\star F_{\alpha\beta} n^\beta$ respectively, on $S$; $n^\mu$
being a unit normal to $\Sigma$. Using the expressions (\ref{em-comp}) for $F_{rt}$
and $F_{\vth\vph}$ given in section \ref{general} and the expressions
(\ref{couplings}) for the couplings $\omega,  \alpha$ and $\beta$, one finds
in the asymptotic limit:
\be  \label{af-charge}
Q_e ~=~  \left(q_e ~-~ q_m b ~\xi_0\right) e^{a \phi_0}~;~~~~~~~~ Q_m ~=~ q_m
\ee
whence
\be \label{em}
F_{tr} ~=~ \frac{\left[Q_e e^{- a \phi_0} - Q_m b
\left(\xi - \xi_0\right)\right] e^{a \phi}}{\left(r - r_0\right)^{2 (1 - n)}
\left(r + r_0\right)^{2 n}} ~dt \wedge dr ~;~~~~~~~~ F_{\vth\vph} ~=~ Q_m \sin \vth ~
d\vth \wedge d\vph.
\ee
In fact it is well-known in the usual string theoretic scenario (corresponding to~ $a = b = 1$) that
a non-vanishing $\xi_0$ is of no consequence by virtue of the Peccei-Quinn (PQ) shift symmetry
of the axion. However, the PQ shift being a classical symmetry of the heterotic action, it can be
broken by instantons in a non-perturbative approach to string theory. We, therefore, retain the
constant $\xi_0$ wherever it is admissible, although it has no significant role in our analysis of
black holes in presence of dilaton and axion.

Using Eqs.(\ref{af-charge}) and the two expressions (\ref{fconsis}a) and (\ref{af-effcoup}) for the
coupling function $\gamma$, we find in the asymptotic limit
\be  \label{af-K1K2}
K_1 ~+~ K_2 ~=~  4 ~q_m^2 ~ e^{- a \phi_0} ~+~ 4 \left(q_e ~-~ q_m b~ \xi_0\right)^2
e^{a \phi_0} ~=~ 4 \left(Q_m^2 ~+~ Q_e^2\right) e^{- a \phi_0} .
\ee

The set of equations (\ref{fconsis}) take the form
\bse
\label{af-eom}
\bea
&& Q_m^2 e^{- a \phi} + \left[Q_e e^{- a \phi_0} - Q_m b \left(\xi - \xi_0\right)\right]^2
e^{a \phi} =~  \frac{Q^2 \left(r - r_0\right)^2 +~ K_2 r r_0}{(r + r_0)^{2 (1 - n)}
(r - r_0)^{2 n}} \label{af-eom-a} \\
&& \phi'^2 ~+~ e^{2 c \phi} \xi'^2 =~ \frac{16 r_0^2 n ~(1 - p)}{(r^2 - r_0^2)^2} \label{af-eom-b}\\
&& 4 Q_m b \left[Q_e e^{- a \phi_0} - Q_m b \left(\xi - \xi_0\right)\right] e^{a \phi} =~
- \frac{\left[\left(r -  r_+\right) \left(r - r_-\right) e^{2 a \phi}~ \xi'\right]'}
{(r + r_0)^{- 2 n} (r - r_0)^{2 (n - 1)}} .  \label{af-eom-c}
\eea
\ese
where $Q^2 \equiv Q_e^2 + Q_m^2$.

In order to determine the constant $r_0$ in terms of the mass $m$, charges $Q_e$ and $Q_m$, we need
to fix atleast one of the parameters $a$ and $b$ so that the Eqs.(\ref{af-eom}) are satisfied.
In the process the values of the other undetermined
parameters $n$ and $K_2$ could also be fixed up. (The constant $K_1$ is of course
given in terms of $Q_m, Q_e, \phi_0$ and $K$ by virtue of Eq.(\ref{af-K1K2})). This is
a very convenient way of deriving complete solutions of the field equations
corresponding to the action (\ref{action-dilax}). We consider
first the special case $|b| = |a|$, which evidently includes the usual string scenario where
$a$ and $b$ are both equal to unity, and then look into the more general choice $|b| \neq
|a|$ and resort to a particular case which is relevant to string theory.

\bigskip
\noindent
{\bf Case ~$|b| = |a|$ :}

\bigskip
\noindent
In this case the set of equations (\ref{af-eom}) can be satisfied uniquely for
$n = 1/(1 + a^2)$ and $K_2 = 0$, whence referring to the expressions (\ref{af-rpm}) one finds
\be  \label{af1-r0}
r_0 ~=~ \frac{(1 + a^2) ~Q^2~ e^{- a \phi_0}}{4 m_0}~;~~ m_0 ~=~ m ~-~
\left(\frac{1 - a^2}{1 + a^2}\right) r_0 ~,~~~~~ \left(Q^2 = Q_e^2 + Q_m^2\right)
\ee
and 
\be
r_+ ~=~ 2 m_0 ~-~ r_0 ~,~~ r_- ~=~ r_0 .
\ee
The metric components $g_{tt} ~(= g^{rr})$ and $g_{\vth\vth} ~(= g_{\vph
\vph}/\sin^2 \vth)$ then take the form
\be
g_{tt} (r) ~=~ - \left(\frac{r + r_0 - 2 m_0}{r + r_0}\right) \left(\frac{r - r_0}{r + r_0}\right)^{\frac{1 - a^2}
{1 + a^2}} ~, ~~~~~ g_{\vth\vth} (r) ~= \left(r + r_0\right)^2 \left(\frac{r - r_0}{r + r_0}\right)^{\frac{2 a^2}
{1 + a^2}}
\ee
By making a coordinate shift ~$r + r_0 \rightarrow r$, the whole set of solutions
can finally be expressed as
\bea \label{af1-metric}
ds^2 ~= &-& \left(1 - \frac{2 m_0} r\right) \left(1 - \frac{2 r_0} r\right)^{\frac{1 - a^2}{1 + a^2}}
dt^2 \nno\\
&+& \left(1 - \frac{2 m_0} r\right)^{- 1} \left(1 - \frac{2 r_0} r\right)^{\frac{a^2 - 1}{a^2 + 1}} dr^2
~+~ r^2 \left(1 - \frac{2 r_0} r\right)^{\frac{2 a^2}{1 + a^2}} d \Omega^2
\eea
\bse
\label{af1-dilax}
\bea
\phi (r) &=& \phi_0 + \frac 1 a  \ln \left[\frac 1 {Q^2} \left(1 - \frac{2 r_0} r\right)^{\frac{- 2 a^2}
{1 + a^2}} \left\{Q_m^2 ~+~ Q_e^2 \left(1 - \frac{2 r_0} r\right)^{\frac{4 a^2}
{1 + a^2}}\right\}\right] \label{af1-dilax-a}\\
\xi (r) &=& \xi_0 + \frac{Q_m Q_e} a e^{- a \phi_0} \left[1 ~-  \left(1 - \frac{2 r_0} r\right)^{\frac{4 a^2}
{1 + a^2}}\right] \left[Q_m^2 ~+~ Q_e^2 \left(1 - \frac{2 r_0} r\right)^{\frac{4 a^2}
{1 + a^2}}\right]^{- 1} \label{af1-dilax-b}
\eea
\ese
\be \label{af1-emsol}
F_{\mu \nu} :~~~~ F_{tr} ~=~ \frac{Q_e}{r^2} ~dt \wedge dr ~;~~~~ F_{\vth\vph} ~=~
Q_m \sin \vth ~d\vth \wedge d\vph
\ee
where $r_0$ and $m_0$ are given in Eqs.(\ref{af1-r0}).

These solutions represent a regular black hole horizon at $r = 2 m_0$ and a curvature singularity
at $r = 2 r_0$. One may now define the total (bare) dilaton and axion charges ($Q_\phi$ and $Q_\xi$
respectively) in a way similar to the total (bare) electromagnetic charges $Q_e$ and $Q_m$ in an
asymptotically flat spacetime:
\bea \label{af1-dilax-charge}
Q_\phi &=& - \frac 1 {4 \pi} \int_S ~^*d \phi ~=~ \frac{a \left(Q_m^2 - Q_e^2\right) e^{- a \phi_0}}
{m_0} \nno\\
Q_\xi &=& - \frac 1 {4 \pi} \int_S ~^*d \xi ~=~ \frac{2 a~ Q_e Q_m}{m_0} ~e^{- 2 a \phi_0} .
\eea
For either $Q_e$ or $Q_m$ equal to zero the axion charge vanishes, while the dilaton charge
flips a sign. This follows from the fact that  a magnetically charged ($Q_e = 0, Q_m = Q$) dilaton black hole transforms to
an electrically charged ($Q_m = 0, Q_e = Q$) dilaton black hole, or vice versa. In effect, this is
a generalization of electric-magnetic duality symmetry in presence of
the dilaton field \cite{gib}-\cite{sen}.

\bigskip
A few special/limiting cases are in order:

\bigskip
\noindent
{\bf (i) $a = 1$:}~ As mentioned earlier, this case corresponds to the bosonic sector of the ten
dimensional heterotic superstring toroidally compactified to four spacetime dimensions.
The above solutions for the spacetime line element and dilaton and axion reduce respectively to
\bea \label{af1-sp1-metric}
ds^2 ~=~ - \left(1 - \frac{2 m} r\right) dt^2 ~+~ \left(1 - \frac{2 m} r\right)^{- 1} dr^2 ~+~
r \left(r - 2 r_0\right) d \Omega^2
\eea
\bea \label{af1-sp1-dilax}
\phi (r) = \phi_0 + \ln \left[\frac{Q^2 r^2 - 4 Q_e^2 r_0 \left(r - r_0\right)}
{Q^2 r \left(r - 2 r_0\right)}\right] ~;~
\xi (r) = \xi_0 + \left[\frac{4 Q_m Q_e e^{- \phi_0} r_0 \left(r - r_0\right)}
{Q^2 r^2 - 4 Q_e^2 r_0 \left(r - r_0\right)}\right]
\eea
where $r_0 = Q^2 e^{- \phi_0}/(2 m)$. The horizon is located at $r = 2 m$ and there is a curvature
singularity at $r = 2 r_0$.

For $Q_e = 0, Q_m = Q$ (or, $Q_m = 0, Q_e = Q$), the solutions correspond to the magnetically
(or, electrically) charged dilaton black hole found independently by Garfinkle, Horowitz and Strominger
(GHS) \cite{ghs} and by Gibbons \cite{gib} (which has also been explained in the works of Gibbons
and Maeda (GM) \cite{gibm}). Accordingly, such a black hole is referred to as a GHS black hole, or
sometimes a GMGHS black hole. Whereas a number of characteristically different dilaton black hole
solutions in arbitrary spacetime dimensions and for arbitrary values of the parameter $a$ have been
illustrated by GM \cite{gibm}, GHS \cite{ghs} elegantly obtained a singly charged (i.e., for either $Q_e
= 0$ or $Q_m = 0$) dilaton black hole in the string case $a = 1$ only. However, the works of both GHS
and GM have been based on the presumption that the KR axion field is zero (or, atleast trivial), which
of course follows automatically when one considers a singly charged configuration. As can
seen readily from the above expressions (\ref{af1-sp1-metric}) and (\ref{af1-sp1-dilax}), the $Q_m = 0,
Q_e = Q$ GMGHS configuration merely produces a mirror image of the solutions corresponding to
$Q_e = 0, Q_m = Q$ under the transformation $\phi \rightarrow - \phi$.

Non-trivial dilaton-axion configurations can be generated from the magnetically (or, electrically) charged
GMGHS dilaton black hole solution (with trivial axion), by utilizing the invariance of the complete set of
field equations under the full symmetry group, viz., the $SL (2,R)$ group \cite{stw,sen}. Such a symmetry
group is generated by repeated applications of the Peccei-Quinn shift of the axion, viz., $\xi \rightarrow
\xi + p$, and the duality $\lambda \rightarrow - 1/\lambda$, where $p$ is a constant and $\lambda \equiv
\xi + i e^{- \phi}$ is a complex scalar constructed out of the dilaton and the axion. The solution generation
process is accomplished on stipulating the gauge field strength to transform under the $SL (2,R)$
transformations as $F_+^{\mu\nu} \rightarrow - \lambda F^{\mu\nu} , ~F_-^{\mu\nu} \rightarrow - \bar{\lambda}
~^\star F^{\mu\nu}$, where $F_\pm^{\mu\nu} = F^{\mu\nu} \pm i ~^\star F^{\mu\nu}$ and $\bar{\lambda}$ is the
complex conjugate of the so-called `axidilaton' field $\lambda$. For $\xi = 0$, the transformation
$\lambda \rightarrow - 1/\lambda$, simply implies $\phi \rightarrow - \phi$. Although the non-trivial dilaton
and axion field configurations which have been found in \cite{stw} by utilizing the duality invariance are not
explicitly in the same form as given above, it can be shown easily that the solutions (\ref{af1-sp1-metric}),
(\ref{af1-sp1-dilax}) indeed result from application of the $SL (2,R)$ transformations on the magnetically (or,
electrically) charged GMGHS dilaton black hole. In fact, it can be shown more rigorously that a non-trivial
dilaton-axion configuration can be generated from the magnetically (or, electrically) charged dilaton black hole
configuration by utilizing the $SL (2,R)$ invariance, even when the parameter $a$ is not necessarily equal to
unity \cite{sspre}.

\bigskip
\noindent
{\bf (ii) $a \ll 1$:}~ For small values of the parameter $a$, the solutions (\ref{af1-dilax}) for the dilaton
and the axion can be expanded as
\be \label{af1-sp2-dilax}
\phi (r) ~=~ \phi_0 + \frac{4 a r_0} r \left(\frac{Q_m^2 - Q_e^2}{Q^2}\right) + {\cal O} (a^3) ~;~~~
\xi (r) ~=~ \xi_0 + \frac{4 a r_0} r \left(\frac{Q_m Q_e}{Q^2}\right) + {\cal O} (a^3)
\ee
where $Q^2 = Q_m^2 + Q_e^2$ and $r_0 = Q^2/(4 m_0) + {\cal O} (a^2)$. Eqs.(\ref{af1-r0}) can be
solved readily to obtain
\be \label{af1-sp2-r0}
r_0 ~=~ \frac 1 2 \left(m ~-~ \sqrt{m^2 - Q^2}\right) +  {\cal O} (a^2) ~,~~~~~ m_0 ~=~ \frac 1 2
\left(m ~-~ \sqrt{m^2 - Q^2}\right) +  {\cal O} (a^2) .
\ee
In the limit $a \rightarrow 0$, Eqs.(\ref{af1-sp2-dilax}) imply that both the dilaton and the axion 
become trivial and one has the familiar Einstein-Maxwell system. The line element (\ref{af1-metric})
takes the form of the standard dyonic Reissner-Nordstr\"om (RN) black hole with horizons at $r = 2 m_0
= m + \sqrt{m^2 - Q^2}$ and at $r = 2 r_0 = m - \sqrt{m^2 - Q^2}$.

\bigskip
\noindent
{\bf (iii) $a \gg 1$:}~ For large values of the parameter $a$, the equations (\ref{af1-r0}) can
be approximated as
\be \label{af1-sp3-r0}
r_0 ~\approx~ \frac{a^2 Q^2 e^{- a \phi_0}}{4 m_0} ~,~~~~~ m_0 ~\approx~ m ~+~ r_0
\ee
Assuming that $\phi_0$ is non-zero and reasonably larger that of $\sim {\cal O} (1/a)$ for large $a$,
one may consider that the constant $r_0$ (and hence $m_0$) does not become arbitrarily larger than
$m$ in the limit $a \rightarrow \infty$. Retaining only the terms upto the leading order in
$1/a$, the solution for the spacetime line element, as well as the solutions for the dilaton $\phi$ and
the axion $\xi$ can approximately be given as
\be  \label{af1-sp3-metric}
ds^2 ~\approx~ - \left(1 - \frac{2 m}{r - 2 r_0}\right) dt^2 ~+ \left(1 - \frac{2 m}{r - 2 r_0}\right)^{- 1} dr^2
~+~ \left(r - 2 r_0\right)^2 d \Omega^2
\ee
\bse
\label{af1-sp3-dilax}
\bea
\phi (r) &\approx& \phi_0 ~+~ \frac 1 a ~\ln \left[\frac 1 {Q^2} \left(1 - \frac{2 r_0} r\right)^{- 2}
\left\{Q_m^2 ~+~ Q_e^2 \left(1 - \frac{2 r_0} r\right)^4\right\}\right] \label{af1-sp3-dilax-a}  \\
\xi (r) &\approx& \xi_0 ~+~ \frac {Q_m Q_e} a~ e^{- a \phi_0} \left[1 -  \left(1 - \frac{2 r_0} r\right)^4\right]
\left[Q_m^2 + Q_e^2 \left(1 - \frac{2 r_0} r\right)^4\right]^{- 1}.    \label{af1-sp3-dilax-b}
\eea
\ese

Of course, one may also neglect the second term in the expression for $\xi$ which contains a suppression
factor of ~$(1/a) ~e^{- a \phi_0}$~ with respect to $\xi_0$. By making a coordinate transformation $r - 2 r_0
\rightarrow r$, we find that the resulting approximate expression for the line element reduces to the standard
Schwarzschild one. However the dilaton, axion and the $U(1)$ gauge field have non-vanishing solutions given by,
\bea
\phi (r) \approx \phi_0 + \frac 1 a \ln \left[\frac{Q_m^2}{Q^2} \left(1 - \frac{2 r_0} r\right)^2
+~ \frac{Q_e^2}{Q^2} \left(1 - \frac{2 r_0} r\right)^{- 2}\right];  \xi (r) = \xi_0 +
{\cal O} \left(\frac{e^{- a \phi_0}} a\right)\\
F_{\mu \nu} :~~~~~~~ F_{tr} ~=~ \frac{Q_e}{r^2} ~dt \wedge dr ~;~~~~ F_{\vth\vph} ~=~
Q_m \sin \vth ~d\vth \wedge d\vph~~~~~~~~~~~~~~~~~
\eea
Solving Eqs.(\ref{af1-sp3-r0}) one finds two possible values of $r_0$ given by $- \frac 1 2
\left(m \pm \sqrt{m^2 + a^2 Q^2 e^{- a \phi_0}}\right)$ which imply two possible limiting
configurations of the dilaton (and also the axion) for large $a$. In the limit $a \rightarrow
\infty$, we have the Schwarzschild line element with both dilaton and axion fields becoming
trivial. This is of course quite expected since under the presumption that for large $a$,
$\phi_0$ is reasonably larger that of $\sim {\cal O} (1/a)$, so that the $e^{- a \phi} F^2$
coupling term in the action virtually disappears. What remains is the coupling $a \xi F ~^\star F$
in the action, and $\xi$ being trivial (as is $\phi$) in the limit $a \rightarrow \infty$,
the $F ~^\star F$ term as usual does not produce any change to the vacuum solution (which is the
Schwarzschild solution) of the field equations corresponding to the pure general relativistic
action\footnote{It is well-known that the addition of a $\int d^4 x \sqrt{- g} F_{\mu\nu}
~^\star F^{\mu\nu}$ term to the pure Einstein action $\int d^4 x \sqrt{- g} R$ does not alter the
vacuum Einstein's equations $R_{\mu\nu} = 0$ and hence the Schwarzschild solution}.

\bigskip
\noindent
{\bf Case $|b| \neq |a|$ :}

\bigskip
\noindent
When the parameters $b$ and $a$ are unequal, it is not always
possible to obtain an analytic closed form black hole solution following the our metric ansatz.
However, some specific choices of values of $b$ and $a$ (where $|b| \neq |a|$), enable one
to fix up the values of the three undetermined constant parameters $r_0, n$ and $K_2$ in
terms of the physical constants $m, Q_e, Q_m$ and $\phi_0$ (and may also $\xi_0$) so that all the
field equations (\ref{af-eom}) are satisfied. We consider the case $b \ll 1$ and $a = 1$, which
is particularly relevant for string theory and can be motivated from two different viewpoints:
\begin{itemize}
\item Firstly, the action in this case closely resembles to the usual dimensionally
reduced action (\ref{het-action}) for the ten dimensional heterotic string apart from
the Planck mass suppressed gauge Chern-Simons term  appearing in the definition of
the modified Kalb-Ramond (KR) three form $H_{\mu\nu\lambda}$ (see Eq.(\ref{axion})).
This term can be neglected in the leading order which amounts to neglecting the Planck
mass suppressed coupling term $\xi F ~^\star F$ that appears in the general action
(\ref{action-dilax}) by virtue of the KR-electromagnetic interaction via the
Chern-Simons term. This, in turn, implies that we are considering the axion field
$\xi$ to be source-free.
\item Secondly, the choice of the parameters $a = 1, b \ll 1$, has a direct correspondence
with an exact low energy string setting (retaining the Chern-Simons term) which contains
a number of unbroken $U (1)$ subgroups belonging to a large gauge group. Such a scenario
has already been discussed in detail in \cite{ortin,kalortin}.
\end{itemize}
For $b \ll 1, a = 1$, Eq.(\ref{af-eom}c) yields the general solution
for the axion in the form of a quadrature:
\be \label{af2-axion}
\xi (r) ~=~ \xi_0 ~+~ \xi_1 \int^r \frac{e^{ - 2 \phi (r')} d r'}{(r' - r_+)(r' - r_-)} ~+~
{\cal O} (b)
\ee
where $\xi_0$ and $\xi_1$ are constants. However, a closer inspection reveals that exact solutions
of Eqs.(\ref{af-eom}a) and (\ref{af-eom}b) determines the metric as well as the dilaton
only when $\xi_1 = 0$. This implies that the axion field $\xi$ is trivial upto ${\cal O} (b)$. In
other words, a complete solution of the full set of equations (\ref{af-eom}) are guaranteed only
when the axion has a negligible effect in shaping the spacetime geometry. Now, the solutions for
the line element and the dilaton $\phi$ correspond exactly to the charged dyonic black hole
\cite{gib,gibm} in four dimensional low energy effective string theory, if the electromagnetic charges
$Q_e$ and $Q_m$ (or, $q_e$ and $q_m$) are associated with two different $U (1)$ gauge fields with
field strengths $F_{\mu\nu}^i ,~ (i = 1, 2)$ \cite{strrev,gibm,ortin,kalortin,agnes}. In such case the axion
$\xi$ becomes automatically source-free and as such the axion charge $Q_\xi$ vanishes. The
source-free property of the axion is retained in a further generalization to the scenario in which
either $U (1)$ fields are due to both electric and magnetic monopole configurations (with respective
charges $Q_e^i$ and $Q_m^i ,~ (i = 1, 2)$) \cite{ortin}. If there are $N$ number of electric and magnetic
charges $Q_e^i$ and $Q_m^i$ ~($i = 1, \cdots, N$) associated with N different $U (1)$ gauge fields
$F_{\mu\nu}^i  ~(i = 1, \cdots, N)$, then it has been shown that \cite{kalortin} an exact dyonic
black hole solution exists only when $\sum Q_e^i Q_m^i = 0$, which in turn implies that
the term $\sum \xi F^i ~^\star F^i$ in the effective action is no longer relevant.
Incidentally, the $\xi$ field turns out to be trivial
and the solution is similar to the dyonic dilaton black hole described in \cite{gib,gibm}.

Neglecting the terms of ${\cal O} (b)$ and considering $\xi_1 = 0$ in the above equation (\ref{af2-axion}),
we write down the full set of solutions of Eqs.(\ref{af-eom}). The solutions are given for ~$n = 1/2$ and
$K_2 = 4 Q_m^2 e^{- \phi_0}$~ (or $K_2 = 4 Q_e^2 e^{- \phi_0}$ --- the solutions are invariant under the
interchange $Q_e \leftrightarrow Q_m$):
\bea \label{af2-metric}
ds^2 ~=~ - \frac{\left(r - r_+\right) \left(r - r_-\right)}{r^2 - r_0^2} ~dt^2 ~+~
\frac{r^2 - r_0^2}{\left(r - r_+\right) \left(r - r_-\right)} ~dr^2 ~+~ \left(r^2 - r_0^2\right)
d\Omega^2
\eea
\bea \label{af2-dilax}
\phi (r) ~=~ \phi_0 ~+~ \ln \left(\frac{r - r_0}{r + r_0}\right) ~;~~~~~~~~ \xi (r) ~=~ \xi_0
\eea
\bea \label{af2-emsol}
&&F_{\mu \nu} :~~~~ F_{tr} ~=~ \frac{Q_e}{(r + r_0)^2} ~dt \wedge dr~;~~~~ F_{\vth\vph} ~=~
Q_m \sin \vth ~d\vth \wedge d\vph
\eea
where
\bea \label{af2-rpm}
r_0 ~=~ \frac{\left(Q_e^2 - Q_m^2\right) e^{- \phi_0}}{2 m} ~; ~~~ \hbox{and} ~~~
r_\pm ~=~ m ~\pm~ \sqrt{m^2 ~+~ r_0^2 ~-  \left(Q_e^2 + Q_m^2\right) e^{- \phi_0}} .
\eea

There are regular horizons located at $r = r_\pm$ and a curvature singularity appears at $r = r_0$. The
dilaton charge turns out to be twice the constant $r_0$ in magnitude:
\be \label{af2-dil-charge}
Q_\phi ~=~ \frac{\left(Q_m^2 ~-~ Q_e^2\right) e^{- \phi_0}} m .
\ee

For either $Q_e$ or $Q_m$ equal to zero, the above solutions reduce respectively to the GHS
magnetically or electrically charged black hole (of course, under the coordinate shift $r + r_0
\rightarrow r$), while in the special cases $Q_e = + Q_m$ or $- Q_m$, the dilaton charge
vanishes and one gets back the standard Reissner-Nordstr\"om black hole.

\subsection{Asymptotically non-flat black holes        \label{dilax-ac}}

We recall the full set of general expressions derived in section \ref{acsol} for
the asymptotically non-flat (AN) spacetime metric, the effective scalar $\psi$, the
effective coupling $\gamma$, as well as the expressions
for the electromagnetic field components, obtained in section \ref{general}. We
summarize below the essential results:

\vskip .5in
\noindent
{\bf Line element:}
\bea \label{ac-line}
ds^2 ~=~ - \frac{\left(r - r_+\right)\left(r - r_-\right)}{r^2 \left(2 r_0/r\right)^{2 n}} ~dt^2
~+~ \frac{r^2 \left(2 r_0/r\right)^{2 n}}{\left(r - r_+\right)\left(r - r_-\right)}~ dr^2 ~+~ r^2
\left(\frac{2 r_0} r\right)^{2 n} d \Omega^2 ~~~~
\eea
{\bf Effective scalar:}
\bea \label{ac-effscalar}
\psi (r) ~=~ \psi_0 ~-~ 2 \sqrt{n (1 - n)} ~\ln \left(\frac {2 r_0} r\right)
\eea
{\bf Effective coupling:}
\bea \label{ac-effcoup}
\gamma (r) ~=~ \left(4 n r^2 ~+~ K_2\right)\left(\frac {2 r_0} r\right)^{2 n}
\eea
{\bf Electromagnetic field components:}
\bea \label{ac-em}
F_{\mu \nu} :~~~~~ F_{tr} ~=~ \frac{\left(q_e ~-~ q_m b ~\xi\right)
e^{a \phi}}{r^2 \left(2 r_0/r\right)^{2 n}} ~dt \wedge dr ~;~~~~~ F_{\vth\vph}
~=~ q_m \sin \vth ~d\vth \wedge d\vph
\eea
where
\bea  \label{ac-rpm}
r_\pm ~=~ \left(\frac 1 {1 - n}\right) \left[m ~\pm~ \sqrt{m^2 ~-~ (1 - n) \frac{K_2} 4}\right] ~.
\eea
The set of equations (\ref{fconsis}) in this case take the form
\bea \label{ac-eom}
(a) && q_m^2 e^{- a \phi} ~+ \left(q_e ~-~ q_m b ~\xi\right)^2 e^{a \phi} ~=~
\left(n r^2 ~+~ \frac{K_2} 4\right) \left(\frac {2 r_0} r\right)^{2 n} \nno\\
(b) && \phi'^2 ~+~ e^{2 a \phi} \xi'^2 ~=~ \frac{4 n (1 - n)}{r^2} \nno\\
(c) && 4 q_m b \left(q_e ~-~ q_m b ~\xi\right) e^{a \phi} ~=~  - r^2 \left(\frac {2 r_0} r\right)^{2 n}
\left[\left(r ~-~ r_+\right) \left(r ~-~ r_-\right) e^{2 a \phi}~ \xi'\right]' .
\eea

We again need to determine the values of the arbitrary constant $K_2$, for which the above
equations (\ref{ac-eom}) are satisfied for appropriate choice of the values of the
parameters $a$ and $b$. The values of the other undetermined parameter $n$ can also be fixed
up in the process. As before, we resort to the special case $|b| = |a|$ first (which includes
the usual string case $b = a = 1$) and then the more general situation $|b| \neq |a|$.
For the latter choice we shall focus to a particular case relevant to string theory.

\bigskip
\noindent
{\bf Case ~$|b| = |a|$}

\bigskip
\noindent
In this case, the solutions of the set of equations (\ref{ac-eom}) are given uniquely for $n =
1/(1 + a^2)$ and $K_2 = 0$, whence referring to the expressions (\ref{ac-rpm}) one finds
$r_+ = 2 m/(1 + n)$ and $r_- = 0$. The full set of solutions can therefore be expressed as
\bea \label{ac1-metric}
ds^2 = - \left(\frac r {2 r_0}\right)^{2 n} \left[1 - \frac {2 m}{(1 - n) r}\right] dt^2 +
\left(\frac {2 r_0} r\right)^{2 n} \left[1 - \frac {2 m}{(1 - n) r}\right]^{- 1} dr^2 +
r^2 \left(\frac {2 r_0} r\right)^{2 n} d\Omega^2 \nno\\
\eea
\bse
\label{ac1-dilax}
\bea
\phi (r) ~=~ \frac 1 a ~\ln \left[\frac{q_m^2 q^4 ~+~ q_e^2 n^2 \left(2 r_0\right)^{4 n} r^{4 (1 - n)}}
{q^4 n \left(2 r_0\right)^{2 n} r^{2 (1 - n)}}\right] \\
\xi (r) ~=~ \frac{q_m q_e} a \left[\frac{q^4 ~-~ n^2 \left(2 r_0\right)^{4 n} r^{4 (1 - n)}}
{q_m^2 q^4 ~+~ q_e^2 n^2 \left(2 r_0\right)^{4 n} r^{4 (1 - n)}}\right]
\eea
\ese
\bea \label{ac1-emsol}
F_{\mu \nu} :~~~~ F_{tr} ~=~ \frac{n q_e}{q^2} ~ dt \wedge dr ~;~~~ F_{\vth\vph} ~=~
q_m \sin \vth ~d\vth \wedge d\vph
\eea
where $q^2 = q_e^2 + q_m^2$.

The constant $r_0$ has dimension of $r$  and the event horizon at $r = 2 m/(1 - n)$ is regular, i.e.,
all the scalar invariants, viz., Ricci, squared Ricci and Kretschmann, are well behaved on that surface;
$r = 0$ of course is a physical singularity. The causal structure is similar to the standard Schwarzschild
black hole and $r = 2 m/(1 - n)$ is a null hypersurface. However, as $r \rightarrow \infty$, the metric
as well as the dilaton and axion fields $\phi$ and $\xi$ are divergent, thus showing the asymptotically
non-flat nature of the spacetime. If the values of either $q_e$ or $q_m$ changes, the metric remains
invariant since it does not involve the gauge charges explicitly.
This also implies that the extremal limit is absolutely redundant for the above line element. The
solutions for the dilaton and the axion, however, depend on $q_e$ and $q_m$ and vary with
either of them. The axion in fact vanishes for either of $q_e$ or $q_m$ equals to  zero, and also for
$r = (q/\sqrt{n})^{1/(1 - n)} (2 r_0)^{- n/(1 - n)}$ even when both the gauge charges are non-vanishing.

Although, the metric above is independent of the electromagnetic charges $q_e$ and $q_m$,
it may still be looked upon as the one that represents a charged black hole. This is because,
not only the metric signifies a dilaton-axion coupled gravito-electromagnetic field configuration,
but also the fact that in the usual string theoretic scenario ($a = 1, n = 1/2$) the string frame
metric $\tilde{g}_{\mu\nu}$ which is related to the Einstein frame metric $g_{\mu\nu}$ via the
conformal transformation $\tilde{g}_{\mu\nu} = e^\phi g_{\mu\nu}$, indeed involves $q_e$ and
$q_m$ through the solution of the conformal factor $e^{\phi}$. In the string frame, the string
length scale $l_s$ is taken as the fundamental unit for all physical measurements, while the
string coupling $g_s$ is determined by the dilaton through the relation $g_s^2 = e^\phi$. Thus
the effective Planck length, i.e., the coefficient of the string frame scalar curvature, varies with
the dilaton. As such, there is no usual way to realize any experimental data physically in the
string frame since the Planck length has a direct bearing upon macroscopic physics through
the strength of gravity. The difficulty is overcome in the Einstein frame in which the Planck
length being a universal constant, is used as a fundamental unit. This is the reason why the
Einstein frame is always preferred and we have concentrated on deriving our solutions in this
frame rather than in the string frame for most part of the present paper. For a recent
discussion on the comparative studies in string and Einstein frames, see \cite{dan}.

In the present case, however, we shall also consider the string frame results (see below in the
discussion for the special case $a = 1$). The solutions for the spacetime line element as well
the dilaton and the axion in a few special/limiting cases for some specific values of the
parameter $a$ are given as follows:

\bigskip
\noindent
{\bf (i) $a = 1$:}~ The above solution (\ref{ac1-metric}) for the (Einstein frame) line element
reduces to
\be \label{ac1-sp1-metric}
ds^2 ~=~ - \left(\frac{r ~-~ 4 m}{2 r_0}\right) dt^2 ~+ \left(\frac{2 r_0}{r ~-~ 4 m}\right) dr^2
~+~ 2 r_0 r~ d \Omega^2
\ee
while the solutions (\ref{ac1-dilax}) for dilaton and axion take the form
\be \label{ac1-sp1-dilax}
\phi (r) ~=~ \ln \left(\frac{q_m^2 q^4 ~+~ q_e^2 r_0^2 r^2}{q^4 r_0 r}\right) ~; ~~~~~~~
\xi (r) ~=~ \frac{q_e q_m \left(q^4 ~-~ r_0^2 r^2\right)}{q_m^2 q^4 ~+~ q_e^2 r_0^2 r^2} .
\ee
The components of the electromagnetic field strength also change: $F_{tr} = q_e/(2 q^2)
dt \wedge dr ; F_{\vth\vph} = q_m \sin \vth d\vth \wedge d\vph$. The event horizon is at
$r = 4 m$ and $r = 0$ is the curvature singularity.

After a coordinate redefinition $\rho = \sqrt{2} r_0 r/q$ the string frame line element as well as
the solutions for dilaton and axion can be expressed in an elegant way as
\be \label{ac1-sp1-stmetric}
d\tilde{s}^2 ~= \left(\frac{q_e^2}{q^2} + \frac{2 q_m^2}{\rho^2}\right)
\left[- \frac{\rho^2}{4 r_0^2} \left(1 - \frac{4 \sqrt{2}~r_0 m}{q \rho}\right) dt^2
~+ \left(1 - \frac{4 \sqrt{2}~ r_0 m}{q \rho}\right)^{- 1} d\rho^2 ~+~ \rho^2
d \Omega^2\right]
\ee
\be \label{ac1-sp1-stdilax}
\phi (r) ~=~ \ln \left(\frac{2 q_m^2 q^2 ~+~ q_e^2 \rho^2}{\sqrt{2} q^3 \rho}\right) ~; ~~~
\xi (r) ~=~ \frac{q_e q_m \left(2 q^2 ~-~ \rho^2\right)}{2 q_m^2 q^2 ~+~ q_e^2 \rho^2}
\ee
where as before $q^2 = q_e^2 + q_m^2$ and $r_0$ now has dimension of square root of the
redefined coordinate $\rho$. The event horizon is at ~$\rho = 4 \sqrt{2} ~r_0 m/q$ and unlike
asymptotically flat spacetimes, the string loop corrections  at large $\rho$ are of much
significance here, as the string coupling $e^\phi \sim \rho$ in the asymptotic limit as long
as $q_e$ is non-zero.

The special cases corresponding to $q_e = 0$ or $q_m = 0$ are of considerable interest.
In either of the instances, the axion field $\xi$ is null and we have magnetically or electrically
charged dilaton black holes with curved asymptotes. There have been detailed studies
on the characteristic properties and thermodynamics of such special (dilaton only)
black hole solutions, found originally by Chan {\it et al} \cite{mann}, in both Einstein and
string frames\footnote{Asymptotically non-flat magnetically charged dilaton black hole
solution for a particular value of the mass $m$ and magnetic charge $q_m$ have also
been obtained in \cite{gershon}.}. However, the general (dyonic) black hole solutions in
either frames, in presence of both non-trivial dilaton and non-trivial axion fields, as shown
respectively in (\ref{ac1-sp1-metric}), (\ref{ac1-sp1-dilax}) and (\ref{ac1-sp1-stmetric}),
(\ref{ac1-sp1-stdilax}) (as well as those given by (\ref{ac1-metric}) - (\ref{ac1-emsol}) in
Einstein frame for arbitrary value of the parameter $a$), are entirely new findings in the present work.

For $q_e = 0$ and $q_m = q$, the solution for the dilaton is $\phi = \ln (q^2/(r_0 r)) =
\ln (\sqrt{2} q/\rho)$. This solution flips a sign for the dual choice $q_m = 0$ and $q_e = q$.
The Einstein metric is obviously invariant since it involves neither $q_e$ nor $q_m$ explicitly. Thus
the electric-magnetic duality, which has been  generalized here in presence of the dilaton, happens to be
a symmetry of nature, as it should be, regardless of the asymptotic properties of the background
spacetime. However, it should be mentioned here that one cannot, in principle, make allusion
to (asymptotic) bare electromagnetic charges $Q_e, Q_m$ in spacetimes that are asymptotically
curved, but rather insist upon the isolated charges $q_e, q_m$. It is also not possible to
define in the usual way a (bare) dilaton charge $Q_\phi$ (nor a bare axion charge $Q_\xi$) in
such spacetimes.

In the string frame, however, the metric certainly changes when one switches over from the
case $q_e = 0, q_m = q$ to the case $q_m = 0, q_e = q$, whence the dilaton $\phi$ flips a
sign. In the latter case, viz, $q_m = 0, q_e = q$, the line element given by
\be \label{ac1-str-elmetric}
d\tilde{s}_e^2 ~= -\frac{\rho^2}{4 r_0^2} \left(1 - \frac{4 \sqrt{2} ~ r_0 m}{q \rho}\right) dt^2
~+ \left(1 - \frac{4 \sqrt{2} ~r_0 m}{q \rho}\right)^{- 1} d\rho^2 ~+~ \rho^2 d \Omega^2
\ee
shows that the horizon is at $\rho = 4 \sqrt{2} ~r_0 m/q$ and the angular part is spherical;
while in the former case, viz, $q_e = 0, q_m = q$, the line element
\be \label{ac1-str-magmetric}
d\tilde{s}_m^2 ~= -\frac{q^2}{2 r_0^2} \left(1 - \frac{4 \sqrt{2} ~r_0 m}{q \rho}\right) dt^2
~+ \frac{2 q^2}{\rho^2} \left(1 - \frac{4 \sqrt{2} ~r_0 m}{q \rho}\right)^{- 1} d\rho^2 ~+~
2 q^2 d \Omega^2
\ee
consists of a cylindrical angular part of radius $\sqrt{2} q$. As pointed out in \cite{mann},
the above metric (\ref{ac1-str-magmetric}) is equivalent to the extremal limit of the
magnetically charged GHS black hole (of course, in the string frame \cite{ghs}) in
coordinates that retain the event horizon.

\bigskip
\noindent
{\bf (ii) $a \ll 1$:}~ For small $a$, $n = 1/(1 + a^2) \approx 1$ and one may approximately
express the above line element (\ref{ac1-metric}) and the solutions (\ref{ac1-dilax}) for
dilaton and axion fields respectively as
\be \label{ac1-sp2-metric}
ds^2 ~\approx~ - \left(\frac r {2 r_0}\right)^2 \left[1 - \frac {2 m}{a^2 r}\right] dt^2 ~+~
\left(\frac {2 r_0} r\right)^{2} \left[1 - \frac {2 m}{a^2 r}\right]^{- 1} dr^2 ~+~ 4 r_0^2 d\Omega^2
\ee
\be \label{ac1-sp2-dilax}
\phi (r) ~\approx~ \frac 1 a ~\ln \left(\frac{q_m^2 q^4 ~+~ 16 q_e^2 r_0^4}{4 q^4 r_0^2}\right) ~;
~~~~~ \xi (r) ~\approx~ \frac{q_m q_e} a \left(\frac{q^4 ~-~ 16 r_0^4}{q_m^2 q^4 ~+~ 16 q_e^2
r_0^4}\right)
\ee
while the electromagnetic field strength components are approximately given by:  $F_{tr}
\approx q_e/q^2 dt \wedge dr ; F_{\vth\vph} = q_m \sin \vth d\vth \wedge d\vph$.
All the above expressions diverge as $a \rightarrow 0$. Therefore the zero limit of $a$ is
ill-defined for the above solutions.

\bigskip
\noindent
{\bf (iii) $a \gg 1$:}~ For large $a$, $n = 1/(1 + a^2) \approx 1/a^2$ and retaining terms
upto ${\cal O} (1/a^2)$, we can approximately write the above line element (\ref{ac1-metric})
and the solutions (\ref{ac1-dilax}) for dilaton and axion fields respectively as
\be \label{ac1-sp3-metric}
ds^2 = - \left(\frac r {2 r_0}\right)^{\frac 2{a^2}} \left[1 - \frac {2 a^2 m}{(a^2 - 1) r}\right] dt^2 +
\left(\frac {2 r_0} r\right)^{\frac 2{a^2}} \left[1 - \frac {2 a^2 m}{(a^2 - 1) r}\right]^{- 1} dr^2 +
r^2 \left(\frac {2 r_0} r\right)^{\frac 2{a^2}} d\Omega^2
\ee
\be \label{ac1-sp3-dilax}
\phi (r) ~\approx~ \frac 2 a ~\ln \left(\frac{a q_m} r\right) ~;
~~~~~ \xi (r) ~\approx~ \frac{q_e}{a q_m} .
\ee
The electromagnetic field strength components are given by: $F_{tr}
\approx q_e/(a^2 q^2) dt \wedge dr ; F_{\vth\vph} = q_m \sin \vth d\vth \wedge d\vph$.

In the limit $a \rightarrow \infty$, the line element looks like the standard Schwarzschild
line element, although the spacetime geometry originally was non-flat asymptotically for finite value of $a$.
This is, however, understandable because of the suppression factor $1/a$ in the
solution for $\phi$. Presence of such a factor implies that the dilaton-electromagnetic coupling term
$e^{- a \phi} F^2$ in the action, is heavily reduced ($\sim 1/a^2$) for very large $a$. This 
automatically ensures a negligible role of both the dilaton and the
electromagnetic field in shaping the spacetime geometry, although the dilaton diverges
logarithmically as $a \rightarrow \infty$. The axion-electromagnetic coupling term $a \xi F
~^\star F$, although present, does not produce any change to the metric solution, since the solution for $\xi$ is
approximately trivial and suppressed by the factor $1/a$. In the limit $a \rightarrow \infty$,
the action in fact reduces to the pure Einstein action supplemented by a ~constant $\times F
~^\star F$ term. In this case, the solution of the Einstein's equations is uniquely given by the Schwarzschild
line element as the term constant $\times F~^\star F$ is a four divergence term in the action and has no significance.
Thus, on the whole, this is a decent check that the Schwarzschild solution indeed
serves as a limit to the dilaton-axion black hole solution (\ref{ac1-metric}) given above, even in
view of the fact that such a solution is presumably non-flat asymptotically.

\bigskip
\noindent
{\bf Case $|b| \neq |a|$ :}

\bigskip
\noindent
As have been the situation for AF spacetimes, here also it is not always possible to obtain
analytic closed form black hole solutions when the parameter $b$ differs in magnitude arbitrarily
from the parameter $a$. We consider a particular case corresponding to typically chosen values
of $a$ and $b$ (of course $|b| \neq |a|$), for which the values of the undetermined constants $n$
and $K_2$ can be ascertained in terms of the physical constants $q_e, q_m$ and $m$ in order
to satisfy the field equations (\ref{af-eom}). The typical choices are $b \ll 1$ and $a = 1$. The
arguments in support of this have already been elucidated in the context of AF black holes in string theory.
Similar to case 2, section \ref{dilax-ac}, here also we find solutions
of the set of equations (\ref{ac-eom}) only when the axion $\xi$ is trivial upto order $b$, i.e., $\xi =
\xi_0 + {\cal O} (b)$. Neglecting the ${\cal O} (b)$ terms, we write down the solutions which
can be shown to be given uniquely for $n = 1/2$ and $K_2 = 2 q_m^2 q_e^2/r_0^2$:
\be \label{ac2-metric}
ds^2 ~=~ - \frac{\left(r ~-~ r_+\right)\left(r ~-~ r_-\right)}{2 r_0 r}~ dt^2 ~+~
\frac{2 r_0 r}{\left(r ~-~ r_+\right)\left(r ~-~ r_-\right)}~ dr^2 ~+~ 2 r_0 r ~d\Omega^2
\ee
\be \label{ac2-dilax}
\phi (r) ~=~ - \ln \left(\frac{q_e^2}{r_0 r}\right) ~;~~~~~~~ \xi (r) ~=~ \xi_0
\ee
\be \label{ac2-emsol}
F_{\mu \nu} :~ F_{tr} ~=~ \frac 1 {2 q_e} ~ dt \wedge dr ~;~~~~~~~
F_{\vth\vph} ~=~ q_m \sin \vth ~d\vth \wedge d\vph
\ee
where
\bea \label{ac2-rpm}
r_\pm ~=~ 2 \left(m ~\pm~ \sqrt{m^2 ~-~ \frac{q_e^2 q_m^2}{4 r_0^2}}\right)  .
\eea
There are two horizons located at $r = r_\pm$, where $r_+$ and $r_-$ are as
given by Eq.(\ref{ac2-rpm}). The horizons are regular, i.e., all the curvature
invariants (i.e., ${\cal R},~ {\cal R}_{\mu\nu} {\cal R}^{\mu\nu}$ and ${\cal R}_{\mu\nu
\alpha\beta} {\cal R}^{\mu\nu\alpha\beta}$) are well behaved in their locations; $r = 0$
continues to give a  physical singularity. The causal structure is similar to the Reissner-Nordstr\"om
black hole. Under the interchange $q_e \rightarrow q_m$, the metric is invariant but
the solution for $\phi$ changes, as does the component $F_{tr}$ of the electromagnetic
field. The essential feature of this class of black hole solutions is that the metric explicitly
involves the electromagnetic charges $q_e$ and $q_m$, which is strikingly different from
the asymptotically non-flat black hole solution (in Einstein frame) encountered in
the case $|b| = |a|$.

\section{Conclusion  \label{conclu}}

It is well known that static spherically symmetric spacetimes incorporating black holes do not
exist in gravity theories minimally coupled to one or more scalar fields. However, the situation
can be remedied either by non-minimal coupling, or the addition of gauge fields, or both. In this
paper, we have considered gravity coupled with dilaton ($\phi$) and Kalb-Ramond axion ($\xi$)
scalar fields and a Maxwell field, arbitrarily coupled to the scalars. The resultant equations of
motion admit of black hole solutions which can be asymptotically flat or asymptotically non-flat.
Generically, the black holes are both electrically and magnetically charged. To admit of the latter,
 the $F_{\mu\nu}~^\star F^{\mu\nu}$ term has been included in the action as well. In
section 2, we considered the most general spherically symmetric metric ansatz and examined
the general properties of the equations of motion for the gravitational, Maxwell and scalar fields.
We found that, in general, the spacetimes that emerge have both electric and magnetic charges,
that $\phi$ and $\xi$ could be combined into a single `effective' scalar field $\psi$, whose
solution contains the constants $r_0$, $k$ and $n$. Among them, the constant
$k$ determines the asymptotic structure of the solution found. In particular, $k=1$
corresponds to asymptotically flat and $k=0$ corresponds to asymptotically non-flat solutions.
For non-trivial $\psi$, one also required the condition: $0 < n < 1$.

In section 3, we found the quasi-local mass of the solutions using well known prescriptions.
For both AF and AN solutions, a reference spacetime had to be `subtracted out' in an
appropriate sense. The resultant quasi-local mass turned out to be positive and finite.

In section 4, we found the location of the event horizon(s) of the black holes as well as the
scalar field $\psi$ in terms of the constants mentioned previously, as well as the electric
and magnetic charges. We also examined the generic large $r$ behaviour of Ricci,
squared-Ricci and Kretschmann scalars and the possible bounds on the mass for the
existence of horizons.

Finally in section 5, we considered a special case of the generic action, inspired by low energy
string theory. The latter fixes the arbitrary coupling of the scalar fields to $F_{\mu\nu}
F^{\mu\nu}$ and $F_{\mu\nu}~^\star F^{\mu\nu}$ terms in the action, upto two
dimensionless constants $a$ and $b$. Such actions have been studied in the context of
string theory before. However, in addition to recovering black hole solutions found earlier, we
get new solutions to the equations of motion. Once again, they can be asymptotically flat or
non-flat. The AF solutions are characterized by the following parameter ranges:
for $a = b$, we obtain a new black hole solution, of which for the special case
$a = b = 1$ we recover the GMGHS solution, while for $a, b \ll 1$, the Reissner-Nordstr\"om solution
is recovered. For $b = 1$ and $a \gg 1$, one essentially arrives at the Schwarzschild solution,
albeit with non-trivial dilaton and axion fields. For $a = 1$ and $b \ll 1$ on the other hand,
we obtain a black hole with two horizons, which is both electrically and magnetically charged,
and of which the Reissner-Nordstr\"om and GMGHS black holes are special cases.
Similarly, for the AN scenario, $a = b$ corresponded to a new black hole solution with both
electric and magnetic charges. For either electric or magnetic charge being zero, we
recover the AN dilaton black hole solution found in \cite{mann}. Finally, for $a = 1$ and $b \ll 1$,
we obtain a dyonic black hole with two horizons.

Thus in this work we present new black hole solutions for
Einstein-Maxwell-scalar field systems inspired by low
energy string theory. It would be interesting to study its generalizations to
rotating black holes. It would also be interesting to study the phenomenological
consequences of these black holes, such as gravitational lensing as well as the
behaviour of particle geodesics, and whether there is any hope of
testing these in astrophysical observations and/or observations in future
high energy colliders (if the higher dimensional Planck energy is indeed small).
Similarly, thermodynamic and stability properties (including quasi-normal
modes) could be of interest. We hope to report on this elsewhere.

\vspace{.5cm}
\noindent
\underline{ACKNOWLEDGMENT}

SD thanks J. Gegenberg for useful discussions. SS and SD thank Indian
Association for the Cultivation of Science, Kolkata, for hospitality, where
part of the work was done. This work is supported in part by the Natural
Sciences and Engineering Research Council of Canada.

\end{document}